\documentclass[aps,prx,twocolumn,showpacs,floatfix,superscriptaddress,longbibliography]{revtex4-2}
\usepackage{graphicx}  
\usepackage{subfigure}  
\usepackage{bm}        
\usepackage{amssymb}   
\usepackage[tbtags]{amsmath}
\usepackage{exscale}
\usepackage{xcolor}
\newcommand{\ket}[1]{|#1\rangle}
\newcommand{\bra}[1]{\langle#1|}

\newcommand{\KM}[1]{{\color{black}{#1}}}

\definecolor{darkred}{rgb}{0.90,0.2,0.2}

\usepackage[plainpages=false,pdfpagelabels,colorlinks=true,linkcolor=red,urlcolor=blue,citecolor=blue,pdftitle={Quantum quenches and relaxation dynamics in the thermodynamic limit},pdfauthor={},pdfdisplaydoctitle=true,pdfduplex=DuplexFlipLongEdge]{hyperref}

\begin{document}

\title{Prethermalization, thermalization, and Fermi's golden rule\\ in quantum many-body systems}

\author{Krishnanand Mallayya}
\affiliation{Department of Physics, Cornell University, Ithaca, New York 14850, USA}
\affiliation{Department of Physics, The Pennsylvania State University, University Park, Pennsylvania 16802, USA}

\author{Marcos Rigol}
\affiliation{Department of Physics, The Pennsylvania State University, University Park, Pennsylvania 16802, USA}

\begin{abstract}
 We study the prethermalization and thermalization dynamics of local observables in weakly perturbed nonintegrable systems, with Hamiltonians of the form $\hat{H}_0+g\hat{V}$, where $\hat{H}_0$ is nonintegrable and $g\hat{V}$ is a perturbation. We explore the dynamics of far from equilibrium initial states in the thermodynamic limit using a numerical linked cluster expansion (NLCE), and in finite systems with periodic boundaries using exact diagonalization. We argue that generic observables exhibit a two-step relaxation process, with a fast prethermal dynamics followed by a slow thermalizing one, only if the perturbation breaks a conserved quantity of $\hat{H}_0$ and if the value of the conserved quantity in the initial state is $\mathcal{O}(1)$ different from the one after thermalization. We show that the slow thermalizing dynamics is characterized by a rate $\propto g^2$, which can be accurately determined using a Fermi golden rule (FGR) equation. \KM{We also show that during such a slow dynamics, observables can be described using projected diagonal and Gibbs ensembles, and we contrast their accuracy.}
\end{abstract}

\maketitle
\section{Introduction}

A far from equilibrium many-body initial state evolving under Hamiltonian dynamics, as relevant to describe an isolated quantum system, is generally expected to relax to thermal equilibrium (in the sense of observables being described by traditional ensembles of statistical mechanics)~\cite{dalessio_kafri_16, Gogolin_2016_Equilibration, Mori_2018_Thermalization}. The advent of quantum simulations in low dimensions with ultracold atoms~\cite{bloch2008many} and trapped ions~\cite{monroe2021programmable} has ushered much of the progress in probing the realm of strongly interacting quantum systems isolated from the environment~\cite{kinoshita2006quantum, trotzky2012probing, gring2012relaxation, langen_erne_15, neill2016ergodic, kaufman_tai_16, Neyenhuise1700672, tang_kao_18, li2020relaxation, malvania2020generalized}. A recurring theme in these experiments~\cite{gring2012relaxation, langen_erne_15, Neyenhuise1700672, tang_kao_18, li2020relaxation} is the phenomenon of prethermalization and its stability to perturbations~\cite{berges2004prethermalization, langen2016prethermalization, ueda2020quantum}.

Prethermalization is a paradigm in which a separation of timescales occur in the relaxation dynamics. In its simplest scenario, two distinct regimes can be identified in the relaxation process: a fast initial dynamics called prethermalization, in which the system relaxes to an intermediate (prethermal) state, followed by a slower relaxation to the true thermal equilibrium (thermalization). These two steps in the dynamics can occur if the system's Hamiltonian has the form $\hat{H}_g=\hat{H}_0+g\hat{V}$, where $\hat{H}_0$ is the so-called reference (or unperturbed) Hamiltonian and $g\hat{V}$ is a perturbation of strength $g$ ($g$ is assumed to be small in a sense to be discussed later). The initial fast evolution is governed by $\hat{H}_0$ and the prethermal state is the equilibrium state of $\hat{H}_0$. The perturbation causes the ensuing slow relaxation to the thermal equilibrium of $\hat{H}_g$, with a relaxation rate that is controlled by the perturbation strength.

However, not every perturbation can lead to a two-step relaxation process because one can split any Hamiltonian into $\hat{H}_0$ and $g\hat{V}$ terms in arbitrary ways, whereas the prethermal and thermal relaxation regimes are qualitatively distinct. Thus one expects $\hat{H}_0$ to have some special feature that uniquely distinguishes it from $\hat{H}_g$. A relatively well understood setup for prethermalization is when $\hat{H}_0$ is integrable and $\hat{V}$ is an integrability breaking perturbation, so that $\hat{H}_g$ is nonintegrable~\cite{moeckel_kehrein_2008, *moeckel_kehrein_2009, *moeckel2010crossover, eckstein_kollar_08, eckstein_kollar_09, kollar_wolf_11, nessi_iucci_14, essler2014quench, essler_fagotti_16, bertini2015prethermalization, *bertini2016prethermalization, fagotti2015universal}. In this case, due to the extensive set of conserved quantities of $\hat{H}_0$, the prethermal state is not a thermal state, but it is described by a generalized Gibbs ensemble (GGE)~\cite{rigol_dunjko_07, cassidy_clark_11, fagotti_2013_reduced, langen_erne_15, vidmar_rigol_16}. The integrability breaking perturbation causes a slow relaxation from the GGE to the thermal equilibrium of $\hat{H}_g$ at a rate $\propto g^2$~\cite{stark_kollar_2013, dalessio_kafri_16, bertini2015prethermalization, *bertini2016prethermalization, mallayya2018quantum, tang_kao_18}.  

A more general setup for prethermalization was discussed in Ref.~\cite{mallayya2019prethermalization}, where it was shown that prethermalization in isolated quantum systems is generic provided the perturbation breaks at least one local (but extensive) conservation law of $\hat{H}_0$ ($\hat{H}_0$ can hence be nonintegrable). Related results in the context of open quantum systems were discussed in Ref.~\cite{lange2018time, lenarvcivc2018perturbative}. In this work, we substantiate the theoretical framework of Ref.~\cite{mallayya2019prethermalization} with further results. We address the following questions: (i) Can one observe a two-step relaxation process for initial states and perturbations that do not change any conserved quantity of $\hat{H}_0$? (ii) How do we calculate the relaxation rates of local observables in the thermodynamic limit? (iii) How do finite-size effects affect the relaxation process?

We carry out our many-body calculations using the numerical linked cluster expansion (NLCE) for dynamics introduced in Ref.~\cite{mallayya2018quantum}, and exact diagonalization (ED) calculations on finite lattices with periodic boundary conditions. The presentation is organized as follows. In Sec.~\ref{sec:setup}, we discuss the general setup for the dynamics, and review earlier results and expectations. In Sec.~\ref{sec:hamnlce}, we introduce the Hamiltonian and observables studied, as well as the computational techniques used. The results of our study are reported in Sec.~\ref{sec:results}, while caveats of studying relaxation rates in finite systems are discussed in Sec.~\ref{sec:finsize}. A summary of our results is presented in Sec.~\ref{sec:summary}.

\section{Set up}\label{sec:setup}

In this section, we describe the quantum quench protocol and review the framework for prethermalization introduced in Ref.~\cite{mallayya2019prethermalization}.

\subsection{Quenches, equilibration, and thermalization} \label{sec_quench}

We study unitary dynamics under a Hamiltonian of the form
\begin{equation}
\hat{H}_g=\hat{H}_0+g\hat{V},
\end{equation} 
where $\hat H_0$ is the reference Hamiltonian and $g\hat{V}$ is a perturbation. The dynamics is initiated through a quantum quench protocol, in which the initial state $\hat{\rho}^{ }_I$ is in thermal equilibrium with respect to an initial Hamiltonian $\hat{H}_I$. The idea is that at time $\tau=0$ the Hamiltonian is suddenly changed from $\hat{H}_I$ to the time-independent final Hamiltonian $\hat{H}_g$. The system  is isolated during the $\tau>0$ dynamics, so that the density matrix evolves unitarily as
\begin{equation}
\hat{\rho}(\tau)=e^{-i\hat{H}_g\tau}\hat{\rho}^{ }_I\,e^{i\hat{H}_g\tau}.\label{eq:rho_tau} 
\end{equation}

We study the dynamics of few-body observables $\hat{O}$, whose expectation values can be written as
\begin{equation}
O(\tau)=\text{Tr}[\hat{O}\hat{\rho}(\tau)].
\end{equation} 
After sufficiently long time evolution, $O(\tau)$ is expected to equilibrate at its steady-state value given by the diagonal ensemble (DE) 
\begin{equation}
O_{\text{DE}}=\text{Tr}[\hat{O}\hat{\rho}^{ }_{\text{DE}}], \quad\text{where}\quad\hat{\rho}^{ }_{\text{DE}}=\lim_{\tau\rightarrow\infty}\frac{1}{\tau}\int_{0}^{\tau}\hat{\rho}(\tau)d\tau\label{O_DE}.
\end{equation}
In the absence of extensive degeneracies in the energy spectrum of $\hat{H}_g$, $\hat{\rho}^{ }_{\text{DE}}$ can be written as~\cite{rigol2008thermalization, dalessio_kafri_16}:
\begin{equation}
\hat{\rho}^{ }_{\text{DE}}=\mathcal{P}^{ }_g[\hat{\rho}^{ }_I]
=\sum_{j}\left(\bra{E_j^g}\hat{\rho}^{ }_I\ket{E_j^g}\right)\ket{E_j^g}\bra{E_j^g}\label{eq:rho_DE},
\end{equation}
where $\{\ket{E_j^g}\}$ are the energy eigenkets of $\hat{H}_g$.

To describe the expectation values of observables after equilibration when $\hat{H}_g$ is nonintegrable, because of eigenstate thermalization~\cite{deutsch1991quantum, srednicki1994chaos, rigol2008thermalization, dalessio_kafri_16}, the diagonal ensemble $\hat{\rho}^{ }_{\text{DE}}$ can be replaced with a Gibbs ensemble $\hat{\rho}^{ }_{\text{GE}}$, 
\begin{equation}
O_{\text{GE}}=\text{Tr}[\hat{O}\hat{\rho}^{ }_{\text{GE}}]\label{O_GE}.
\end{equation}
This can be done because one can show under very general physically relevant assumptions that, as a result of eigenstate thermalization, $O_{\text{DE}}=O_{\text{GE}}$ for few-body or local operators $\hat{O}$~\cite{dalessio_kafri_16}. We stress that $\hat{\rho}^{ }_{\text{GE}}$ needs to be determined taking into account the conserved quantities of $\hat{H}_g$. In this work, we consider a nonintegrable $\hat H_0$ with only one conserved quantity, the number of particles $\hat{N}$. 

For a perturbation that breaks particle-number conservation, the thermal state $\hat{\rho}^{ }_{\text{GE}}$ is given by
\begin{equation}\label{eq:rho_GE}
\hat{\rho}^{ }_{\text{GE}}=  
\dfrac{ e^{-\hat{H}_g/T}}{ \text{Tr}[e^{-\hat{H}_g/T}]},   \quad\text{when}\quad[\hat{H}_g,\hat{N}]\ne0.
\end{equation}
The temperature $T$ is the only parameter that needs to be determined in $\hat{\rho}^{ }_{\text{GE}}$. To find $T$ one uses the conservation of energy under the unitary evolution dictated by $\hat{H}_g$:
\begin{equation}
\text{Tr}[\hat\rho^{ }_{\text{GE}}\hat H_g]=\text{Tr}[\hat \rho^{ }_{I}\hat H_g]\label{finalT}.
\end{equation}

If the perturbation does not break particle-number conservation, then $\hat{\rho}^{ }_{\text{GE}}$ has the form
\begin{equation}\label{eq:rho_GE_N}
\hat{\rho}^{ }_{\text{GE}}=  
\dfrac{ e^{-(\hat{H}_g+\mu\hat{N})/T}}{ \text{Tr}[e^{-(\hat{H}_g+\mu\hat{N})/T}]}, \quad\text{when}\quad [\hat{H}_g,\hat{N}]=0.
\end{equation}
The parameters $T$ and $\mu$ are determined in this case using the conservation of energy~\eqref{finalT} and of the number of particles
\begin{equation}
\text{Tr}[\hat\rho^{ }_{\text{GE}}\hat N]=\text{Tr}[\hat \rho_{I}\hat N]\label{finalmu}.
\end{equation} 

\subsection{Prethermalization and projected dynamics}

Here we summarize the framework of prethermalization and the assumptions involved, as introduced in Ref.~\cite{mallayya2019prethermalization}. The main assumption is a ``weak coupling'' condition $g\tau_0\ll 1$, where $\tau_0$ is the time scale at which equilibration occurs under the reference (unperturbed) dynamics dictated by $\hat{H}_0$. Since we want to avoid having $\tau_0$ and the perturbed dynamics to depend on the system size, something that would happen if there is net transport of particles and/or energy between different parts of the system, we assume that the initial states and the Hamiltonians are translationally invariant. Under the weak coupling assumption, thermalization occurs at times $\tau\gg\tau_0$.

Another condition needed for one to be able to observe a two-step relaxation process is that the initial value of the conserved quantity per site, in our case $n_I=\text{Tr}[\hat{N}\hat{\rho}^{ }_I]/L$ ($L$ is the number of lattice sites), should be sufficiently different from its equilibrated value, $n^{ }_{\text{DE}}=\text{Tr}[\hat{N}\hat{\rho}^{ }_{\text{DE}}]/L$, namely, that $|n_I-n^{ }_{\text{DE}}|\sim\mathcal{O}(1)>\mathcal{O}(g)$. This ensures that the prethermal and thermal results for the observable are separated beyond $\mathcal{O}(g)$ corrections. This condition was not explicitly stated in Ref.~\cite{mallayya2019prethermalization}, and we will justify it with numerical results in Sec.~\ref{Sec_Perturbations}.

Now let us specify what we mean, in the two-step relaxation process, by fast prethermalization followed by a slow thermalization.\\
(i) \textbf{\textit{Fast prethermal dynamics}} at $\tau\lesssim\tau_0$: $O(\tau)$ resembles the reference dynamics and relaxes to a prethermal value. This value is determined by the DE of $\hat{H}_0$, which is $\mathcal{P}^{ }_0[\hat{\rho}^{ }_I]$ [Eq.~\eqref{eq:rho_DE}], with $\mathcal{O}(g)$ corrections.

(ii) \textbf{\textit{Slow thermalization}} at $\tau\gg\tau_0$:  $O(\tau)$ relaxes slowly to the true equilibrium value $O_{\text{DE}}$ [Eq.~\eqref{O_DE}]. The slow relaxation proceeds via intermediate equilibrium states of $\hat{H}_0$. These intermediate equilibrium states are projected diagonal ensembles (P-DEs) of $\hat{H}_0$, $\hat{\rho}^{ }_{\text{P-DE}}(\tau)$. They are obtained acting with the projector $\mathcal{P}^{ }_0$ on $\hat{\rho}(\tau)$,
\begin{eqnarray}
\hat{\rho}^{ }_{\text{P-DE}}(\tau)&=&\mathcal{P}^{ }_0[\hat{\rho}(\tau)]\nonumber\\
&=&\sum_{i}\left(\bra{E_i^0}\hat{\rho}(\tau)\ket{E_i^0}\right)\ket{E_i^0}\bra{E_i^0}\label{proj_DE},
\end{eqnarray}
where $\{\ket{E_i^0}\}$ are the simultaneous eigenkets of $\hat{H}_0$ and $\hat{N}$. $\hat{\rho}^{ }_{\text{P-DE}}(\tau)$ defines the projected dynamics for the observable $\hat{O}$
\begin{equation}
O_{\text{P-DE}}(\tau)=\text{Tr}\left\{\hat{O}\hat{\rho}^{ }_{\text{P-DE}}(\tau)\right\}.\label{eq:O_proj_DE}
\end{equation}
By construction, the exact dynamics and the projected dynamics are identical for the conserved quantities of the reference Hamiltonian, namely, for $\hat{N}$ and $\hat{H}_0$. For other observables $\hat{O}$, an $\mathcal{O}(g)$ discrepancy is expected between $O(\tau)$ and $O_{\text{P-DE}}(\tau)$ for $\tau\gg\tau_0$. The initial value $O_{\text{P-DE}}(\tau=0)$ is that of the prethermal state.

In the thermodynamic limit when $\hat{H}_0$ is nonintegrable, because of eigenstate thermalization, the expectation values of observables can be computed replacing P-DE by a Gibbs ensemble (P-GE):
\begin{equation}
O_{\text{P-GE}}(\tau)=\text{Tr}[\hat O \hat{\rho}^{ }_{\text{P-GE}}(\tau)]\label{eq:O_PGE},
\end{equation}
where $\hat{\rho}^{ }_{\text{P-GE}}(\tau)$ is given by
\begin{equation}
\hat{\rho}^{ }_{\text{P-GE}}(\tau)=\dfrac{ e^{-\left[\hat{H}_0+\mu(\tau)\hat{N}\right]/T(\tau)}}{ \text{Tr}\left\{e^{-\left[\hat{H}_0+\mu(\tau)\hat{N}\right]/T(\tau)}\right\}},\label{eq:rho_Proj_GE}
\end{equation}
with $T(\tau)$ and $\mu(\tau)$ determined by the instantaneous values of the conserved quantities of $\hat{H}_0$:
\begin{eqnarray}
\text{Tr}[\hat{\rho}^{ }_{\text{P-GE}}(\tau)\hat H_0]&=&\text{Tr}[\hat \rho(\tau)\hat H_0]\label{eq:proj_finalT},\\
\text{Tr}[\hat{\rho}^{ }_{\text{P-GE}}(\tau)\hat N]&=&\text{Tr}[\hat \rho(\tau)\hat N]\label{eq:proj_finalmu}.
\end{eqnarray}
The dynamics of the P-GE is thus dictated by the evolution of the broken conserved quantities (per site) of $\hat{H}_0$, namely,  $n(\tau)=\text{Tr}[\hat \rho(\tau)\hat N]/L$ and  $e_0(\tau)=\text{Tr}[\hat \rho(\tau)\hat H_0]/L$. Their dynamics is described by an autonomous equation based on Fermi's golden rule (FGR). For $n(\tau)$, this is given by
\begin{eqnarray}
\dfrac{dn}{d\tau}&=&\dfrac{2\pi g^2}{L}\sum_{i,j} \delta(E^0_j-E^0_i) \left(\bra{E^0_j}\hat N\ket{E^0_j}-\bra{E^0_i}\hat N\ket{E^0_i}\right)\nonumber\\&&\times\left|\bra{E^0_j}\hat{V}\ket{E^0_i}\right|^2P^0_i(\tau) + \mathcal{O}(g^3),\label{fermi_rate}
\end{eqnarray}
where $\{\ket{E_i^0}\}$ are the simultaneous eigenkets of $\hat{H}_0$ and $\hat{N}$, and $P^0_i(\tau)$ are the diagonal matrix elements of the P-DE at $\tau$. $P^0_i(\tau)$ can equivalently be replaced with the diagonal matrix elements of the P-GE, $\bra{E_i^0}\hat{\rho}^{ }_{\text{P-GE}}(\tau)\ket{E_i^0}$, which in turn are determined by $n(\tau)$ and $e_0(\tau)$. In Sec.~\ref{sec_rate_NLCE}, we refine this FGR expression to obtain a relaxation rate. 

The reference energy $e_0(\tau)$ is approximately constant as $\hat{H}_0$ and $\hat{H}_g$ differ only by the perturbation term, and $\hat{H}_g$ is an exact conserved quantity in the dynamics. If energy is not conserved, such as when $\hat{V}$ is time dependent, a similar autonomous equation can describe the evolution of $e_0(\tau)$. In the context of periodically driven perturbations, the FGR based equation was used to compute heating rates in Ref.~\cite{mallayya2019heating}.

(iii) \textbf{\textit{Thermal equilibrium}} at $\tau\rightarrow\infty$:
After sufficiently long time, the system attains its final equilibrium state under $\hat{H}_g$, predicted by the DE [Eq.~\eqref{eq:rho_DE}]. Since we consider a nonintegrable $\hat{H}_g$, $\hat\rho^{ }_{\text{DE}}$ can be replaced by  $\hat\rho^{ }_{\text{GE}}$ [in Eq.~\eqref{eq:rho_GE}] to compute the expectation values of observables after equilibration.

The projected dynamics $O_{\text{P-DE}}(\tau)$ [Eq.~\eqref{proj_DE}] also attains an equilibrium steady state value $\overline{O}_{\text{P-DE}}=\lim_{\tau\rightarrow\infty}\tau^{-1}\int_{0}^{\tau}O_{\text{P-DE}}(\tau)d\tau$ given by
\begin{equation}
\overline{O}_{\text{P-DE}}=\text{Tr}\left\{\hat{O}\mathcal{P}^{ }_0\left[\hat{\rho}^{ }_{\text{DE}}\right]\right\},\label{eq:Obar_PDE}
\end{equation}
where the density matrix $\mathcal{P}^{ }_0\left[\hat{\rho}^{ }_{\text{DE}}\right]$ describes the equilibrium state of the projected dynamics ($\overline{\text{P-DE}}$).  

With nonintegrable $\hat{H}_0$, $\overline{O}_{\text{P-DE}}$ can be replaced with the thermal value $\overline{O}_{\text{P-GE}}$ given by
\begin{equation}
\overline{O}_{\text{P-GE}}=\text{Tr}\left[\hat{\overline{\rho}}^{ }_{\text{P-GE}}\hat{O}\right]\label{eq:O_PGE_bar},
\end{equation}
where $\hat{\overline{\rho}}^{ }_{\text{P-GE}}$ is the Gibbs ensemble describing this thermal equilibrium ($\overline{\text{P-GE}}$). $\hat{\overline{\rho}}^{ }_{\text{P-GE}}$ is given by Eq.~\eqref{eq:rho_Proj_GE} but with $\bar{T}$ and $\bar{\mu}$ instead of $T(\tau)$ and $\mu(\tau)$, respectively, determined by the equations
\begin{eqnarray}
\text{Tr}[\hat{\overline{\rho}}^{ }_{\text{P-GE}}\hat H_0]&=&\text{Tr}[\hat \rho^{ }_{\text{GE}}\hat H_0]\label{eq:proj_finalT_bar},\\
\text{Tr}[\hat{\overline{\rho}}^{ }_{\text{P-GE}}\hat N]&=&\text{Tr}[\hat \rho^{ }_{\text{GE}}\hat N]\label{eq:proj_finalmu_bar},
\end{eqnarray}
where $\hat \rho^{ }_{\text{GE}}$ is the true thermal state of $\hat{H}_g$ [Eq.~\eqref{eq:rho_GE}].

\section{Hamiltonian, observables, and computational techniques}\label{sec:hamnlce}

In this section we introduce the model Hamiltonian, the observables of interest, and the numerical techniques used in our calculations. 

\subsection{Hamiltonians and observables}

We study the dynamics of strongly interacting hard-core bosons in a translationally invariant one-dimensional lattice with $L$ sites. Let $\hat{b}_i^{\dagger}$ $(\hat{b}_i^{})$ be the creation (annihilation) operator of a hard-core boson at site $i$. They satisfy bosonic commutation relations with the hard-core constraint: $\hat{b}_i^2=(\hat{b}_i^{\dagger})^2=0$. 

Our reference Hamiltonian is
\begin{eqnarray}
&&\hat{H_0}=\sum_i \left[ -t\left( \hat{b}^\dagger_i \hat{b}^{}_{i+1} + \textrm{H.c.} \right) -t'\left( \hat{b}^\dagger_i \hat{b}^{}_{i+2} + \textrm{H.c.} \right) 
\right.\label{model_H}\\
&&\left.+V\left(\hat{n}^{}_i-\dfrac{1}{2}\right)\hspace*{-0.1cm}\left(\hat{n}^{}_{i+1}-\dfrac{1}{2}\right)+V'\left(\hat{n}^{}_i-\dfrac{1}{2}\right)\hspace*{-0.1cm}\left(\hat{n}^{}_{i+2}-\dfrac{1}{2}\right)\hspace*{-0.05cm}\right],\nonumber
\end{eqnarray}
where $\hat{n}_i=\hat{b}^{\dagger}_{i}\hat{b}^{ }_{i}$, $t$ ($t'$) is the nearest (next-nearest)-neighbor hopping and $V$ ($V'$) is the nearest (next nearest)-neighbor interaction strength. We always consider $t\ne0$ and $V\ne0$, and fix $t'=V'=0.7$, so that $\hat{H}_0$ is nonintegrable~\cite{santos2010onset}. The total number of particles, $\hat{N}=\sum_i \hat{b}_i^{\dagger}\hat{b}_i^{}$, is the only local conserved quantity of this Hamiltonian, $[\hat{H}_0,\hat{N}]=0$. We consider two perturbations of $\hat{H}_0$, which we denote as $g_1\hat{V}_1$ and $g_N\hat{V}_N$.

Our first perturbed Hamiltonian has the form $\hat{H}_{g_1}=\hat{H}_0+g_1\hat{V}_1$, where
\begin{equation}
g_1\hat{V}_1 = g_1\sum_i\left[\hat{b}^{}_{i}+\frac{1}{2}\left(\hat{b}^{}_{i}\hat{b}^{}_{i+1} \right)+ \text{H.c.}\right].\label{H1}
\end{equation}
We note that $\hat{H}_{g_1}$ is not particle-number conserving, because $[\hat{V}_1,\hat{N}]\ne0$, but it is particle-hole symmetric.

Our second perturbed Hamiltonian has the form $\hat{H}_{g_N}=\hat{H}_0+g_N\hat{V}_N$, where
\begin{equation}
g_N\hat{V}_N = g_N\sum_i\left[\dfrac{1}{2}\left(\hat{b}^{\dagger}_{i}\hat{n}_{i+1}\hat{b}^{ }_{i+2}\right)+ \text{H.c.}\right].\label{eq:HN}
\end{equation}
This perturbed Hamiltonian is particle-number conserving, because $[\hat{V}_N,\hat{N}]=0$, but it is not particle-hole symmetric, because $\hat{V}_N$ breaks that symmetry of $\hat{H}_0$. In what follows we use $\hat{H}_g=\hat{H}_0+g\hat V$ as a common notation for $\hat{H}_{g_1}=\hat H_0+g_1\hat V_1$ and $\hat{H}_{g_N}=\hat H_0+g_N\hat V_N$. We only make a distinction between the two Hamiltonians when needed. 

The pre-quench Hamiltonian $\hat{H}_I$, which determines the initial state $\hat{\rho}^{ }_I$, is also described by Eq.~\eqref{model_H} but with different coupling parameters $t_I$ and $V_I$. This ensures that $[\hat H_I,\hat H_0]\ne 0$ so that $\hat{\rho}^{ }_I$ has nontrivial reference dynamics. $\hat{\rho}^{ }_I$ is in thermal equilibrium with respect to $\hat{H}_I$ at a temperature $T_I$ and a chemical potential $\mu_I$:
\begin{equation}
\hat{\rho}^{ }_I=\dfrac{ e^{-(\hat{H}_I+\mu_I\hat{N})/T_I}}{ \text{Tr}[e^{-(\hat{H}_I+\mu_I\hat{N})/T_I}]}.\label{eq:rho_I}
\end{equation}
$\mu_I$ allows us to control the site occupation of the initial state, $n(\tau=0)=\text{Tr}[\hat{N}\hat{\rho}^{ }_I]/L$. Due to the particle-hole symmetry of $\hat{H}_I$, $\mu_I=0$ corresponds to half filling. 

\textit{Parameters for the numerical calculations}: We take $\hat{H}_0$ [Eq.~\eqref{model_H}] to have $t=V=1$, so that both nearest- and next-nearest-neighbor terms make $\hat{H}_0$ nonintegrable. For the pre-quench Hamiltonian $\hat{H}_I$, we take $t_I=0.5$ and $V_I=1.5$ so that the quench is not small and the system is far from equilibrium at time $\tau=0$. The numerical results are generic and not sensitive to the choice of these coupling parameters. For the initial state $\hat{\rho}^{ }_I$ [Eq.~\eqref{eq:rho_I}], we take $T_I=10$. We consider the initial chemical potential $\mu_I=0$ to study quenches at half filling, and $\mu_I=1.5$ to study quenches away from half filling (so that the initial site occupation is $n_I= 0.47$). 

\textit{Observables:} We study the dynamics of two extensive observables: \\
(i) The total number of particles $\hat{N}$, whose expectation value per site (the site occupation) has been denoted as $n$. $\hat{N}$ is the only local conserved quantity of $\hat{H}_0$. \\
(ii) The nearest neighbor density correlator 
\begin{equation}
\hat{U}=\sum_i\left(\hat{n}^{}_i-\dfrac{1}{2}\right)\hspace*{-0.1cm}\left(\hat{n}^{}_{i+1}-\dfrac{1}{2}\right),\label{eq:U}
\end{equation}  
whose expectation value per site is denoted as $u$. This is an experimentally accessible local observable that exhibits nontrivial dynamics under both $\hat{H}_0$ and $\hat{H}_g$. The dynamics of other local observables such as the nearest-neighbor one-body correlator $\left(\sum_i \hat{b}^\dagger_i \hat{b}^{}_{i+1} + \textrm{H.c.}\right)$ are qualitatively similar to that of $u(\tau)$.

\subsection{Computational approaches}

\subsubsection{Numerical linked cluster expansion (NLCE)}

We use a numerical linked cluster expansion (NLCE) to calculate the expectation value of extensive observables $\hat{O}$ per site, $\langle \hat{O}\rangle/L$, in the thermodynamic limit~\cite{rigol2006numerical, *rigol2007numerical1, *rigol2007numerical2}. NLCE allows one to compute $\langle \hat{O}\rangle/L$ as a sum over contributions from all connected clusters $c$ that can be embedded on the lattice:
\begin{equation}\label{nlce_eq}
\langle \hat{O}\rangle/L=\sum_{c}M(c)\times W_{O}(c),
\end{equation}
where $W_{O}(c)$ is the weight of cluster $c$, and $M(c)$ is the number of ways per site to embed the cluster $c$ in the lattice. $W_O(c)$ is computed for each cluster $c$ using the inclusion-exclusion principle:
\begin{equation}\label{weight_subtraction}
W_{O}(c)=\langle\hat{O}\rangle_c- \sum_{s \subset c} W_{O}(s),
\end{equation}
where $\langle\hat{O}\rangle_c$ is the expectation value of $\hat{O}$ in cluster $c$ and $s\subset c$ denotes all connected sub-clusters of $c$. For the smallest cluster $c_0$, $W_{O}(c_0)=\langle\hat{O}\rangle_{c_0}$. 

For each cluster $c$, $\langle\hat{O}\rangle_c = \text{Tr}[\hat{\rho}^c\hat{O}]$, where $\hat{\rho}^c$ is the relevant density matrix in the cluster. The appropriate cluster Hamiltonians $\hat{H}^c$ that define $\hat{\rho}^c$ are modified from their definition in the thermodynamic limit to fit the sites and bonds in the cluster. For example, the initial state, $\hat{\rho}^{c}_I$ is
given by Eq.~\eqref{eq:rho_I} with the Hamiltonian $\hat{H}_I\rightarrow\hat{H}^{c}_I$ for the cluster. Similarly, $\hat{\rho}^c(\tau)$ [Eq.~\eqref{eq:rho_tau}],  $\hat{\rho}^{c}_{\text{DE}}$ [Eq.~\eqref{eq:rho_DE}],  $\hat{\rho}^{c}_{\text{GE}}$  [Eqs.~\eqref{eq:rho_GE} and~\eqref{eq:rho_GE_N}], $\hat{\rho}^{c}_{\text{P-DE}}(\tau)$ [Eq.~\eqref{proj_DE}] and $\hat{\rho}^{c}_{\text{P-GE}}(\tau)$ [Eq.~\eqref{eq:rho_Proj_GE}] are evaluated using the cluster modified Hamiltonians $\hat{H}_g\rightarrow\hat{H}^c_g$ and $\hat{H}_0\rightarrow\hat{H}^c_0$. $\langle\hat{O}\rangle_c$ is calculated numerically using full exact diagonalization. 

Since our model has nearest- and next-nearest-neighbor bonds, one has the freedom to choose different building blocks to construct the clusters in NLCE~\cite{mallayya2017numerical}. We use the maximally connected expansion~\cite{rigol2014quantum}, in which each cluster $c$ is made of contiguous sites with all possible bonds present in the cluster modified Hamiltonians. The maximally connected expansion is ideal for dynamics~\cite{mallayya2018quantum}, as well as for diagonal and grand canonical ensemble calculations, in our one-dimensional lattices~\cite{rigol2014quantum, mallayya2017numerical}.

The order of the NLCE is set by the number of sites of the largest cluster used in the expansion. We compute the NLCE up to the 19$^{\text{th}}$ order for quenches that conserve $\hat{N}$, after exploiting reflection symmetry in the particle number sectors (the dimension of the largest sector is 46252). For quenches that break particle conservation, we exploit particle-hole and reflection symmetry to compute the NLCE up to the 18$^{\text{th}}$ order (the largest sector dimension is 65792).

\subsubsection{Exact diagonalization (ED)}

For the quenches that break particle-number conservation, we also study dynamics in finite lattices with $L$ sites and periodic boundary conditions (PBC) solved using ED. The largest chain we solve has $L=19$ sites (whose largest sector dimension is 13797).

We also use ED to compute the FGR rate in Eq.~\eqref{eq:FGR_rate_N} [see Sec.~\ref{sec_rate_NLCE} and Appendix.~\ref{appd}]. Evaluating Eq.~\eqref{eq:FGR_rate_N} does not involve dynamics, and requires only the diagonalization of $\hat{H}_0$, which conserves $\hat{N}$.  We evaluate Eq.~\eqref{eq:FGR_rate_N} in chains with $L=22$ sites (whose largest sector dimension is 32065).

\section{Results}\label{sec:results}

\subsection{Perturbations and initial states}\label{Sec_Perturbations}

In Ref.~\cite{mallayya2019prethermalization}, we studied prethermalization and thermalization in the context of perturbations that break a conserved quantity, and initial states for which the expectation value of the conserved quantity was different from the value after equilibration. Whether this two-step relaxation process occurs for perturbations or initial states that do not change the value of the conserved quantity is the question that we address next. 

\begin{figure}[!t]
\includegraphics[width=0.985\linewidth]{ 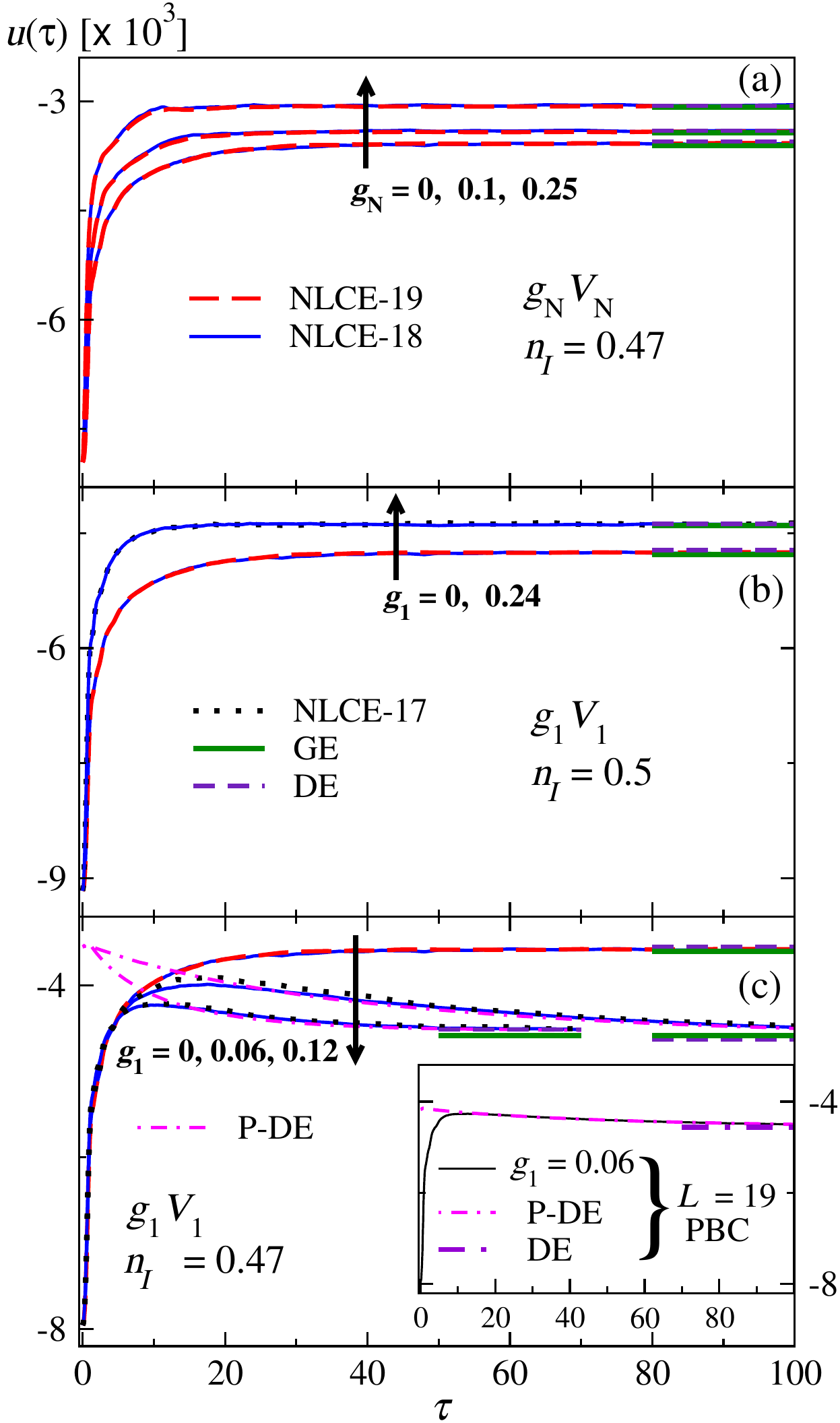}
  \caption{Dynamics of the nearest-neighbor correlation $u(\tau)$ [Eq.~\eqref{eq:U}] after a quench under three scenarios (see text). (a) The perturbation $g_N\hat{V}_N$ conserves the particle number. The initial site occupation is $n_I=0.47$ [set with $\mu_I=1.5$ and $T_I=10$ in Eq.~\eqref{eq:rho_I}]. (b) The perturbation $g_1\hat{V}_1$ breaks particle-number conservation. $n_I=1/2$ ($\mu_I=0$), and $T_I=10$. (c) (main panel and inset) The same perturbation $g_1\hat{V}_1$ as in (b) and the same initial state as in (a) ($n_I=0.47$). All main panels show $u(\tau)$ evaluated at the highest two orders of the NLCE: 19$^{\text{th}}$ order (NLCE-19) and 18$^{\text{th}}$ order (NLCE-18) for $\hat{N}$-commuting Hamiltonians [all values of $g_N$ in (a) and $g_1=0$ in (b) and (c)], and 18$^{\text{th}}$ order and 17$^{\text{th}}$ order (NLCE-17) in all others. The inset in (c) shows $u(\tau)$ in a finite chain with $L=19$ sites and periodic boundary conditions (PBC) solved using ED.  All panels show the final equilibrium values given by the DE (shown for the highest order of the NLCE in the main panels, and $L=19$ sites in the inset) and GE prediction (in the main panels, evaluated to machine precision with NLCE). Also shown in (c) are the projected dynamics (P-DE) results for $g_1>0$, obtained using the 18$^{\text{th}}$-order NLCE (main panel) and $L=19$ sites with ED (inset).}
\label{Fig:Pert_dynamics}
\end{figure}

In Fig.~\ref{Fig:Pert_dynamics} we show results for $u(\tau)$ [Eq.~\eqref{eq:U}] under three different scenarios. In Fig.~\ref{Fig:Pert_dynamics}(a), the system evolves under a particle-number-conserving Hamiltonian, $\hat{H}_{g_N}=\hat{H}_0+g_N\hat{V}_N$ [Eq.~\eqref{eq:HN}]. In Fig.~\ref{Fig:Pert_dynamics}(b), the system evolves under $\hat{H}_{g_1}=\hat{H}_0+g_1\hat{V}_1$ [Eq.~\eqref{H1}], which breaks particle-number conservation, but the initial state is at half filling. Due to the particle-hole symmetry of $\hat{H}_{g_1}$, the system remains at half filling and the expectation value of the broken conserved quantity does not evolve in time. In Fig.~\ref{Fig:Pert_dynamics}(c), the system evolves under $\hat{H}_{g_1}=\hat{H}_0+g_1\hat{V}_1$ [Eq.~\eqref{H1}] as in Fig.~\ref{Fig:Pert_dynamics}(b), but the initial state is away from half filling. After equilibration under the dynamics dictated by $\hat{H}_{g_1}$, the system must be at half filling due to the particle-hole symmetry of $\hat{H}_{g_1}$. Thus the expectation value of the broken conserved quantity evolves in time.

The main panels in Fig.~\ref{Fig:Pert_dynamics} show $u(\tau)$ in the thermodynamic limit as obtained using the NLCE. The fact that curves for the last two orders of the NLCE overlap indicate that convergence errors are small during the dynamics. The equilibrium values predicted by the diagonal ensemble (DE) are evaluated in the highest order of the NLCE, and are shown as horizontal dashed lines. The thermal equilibrium values predicted by the grand canonical ensemble (GE) are also evaluated using the NLCE, and are shown as horizontal solid lines. The GE results converge exponentially faster than the DE ones, and attain machine precision already at much lower orders of the NLCE~\cite{rigol2014quantum, *rigol2016fundamental, mallayya2017numerical}. To obtain the exact value in the GE within machine precision, a 15$^{\text{th}}$-order NLCE is sufficient as the relative convergence errors become $<10^{-12}$. The agreement between the DE and GE results in the main panels in Fig.~\ref{Fig:Pert_dynamics}(a)--\ref{Fig:Pert_dynamics}(c) show that the system attains thermal equilibrium in the three scenarios considered, as expected for nonintegrable systems~\cite{rigol2014quantum, rigol2016fundamental}. The small discrepancy between the DE and GE results is a consequence of the relatively slower convergence of the DE calculations. In Appendix~\ref{appa}, we show that the DE estimate for $\hat{U}$ converges towards the GE result with increasing orders of the NLCE. 

One can see in Fig.~\ref{Fig:Pert_dynamics}(a) and~\ref{Fig:Pert_dynamics}(b) that $u(\tau)$ relaxes to the thermal equilibrium result in a single step. On the other hand, Fig.~\ref{Fig:Pert_dynamics}(c) shows the characteristic two-step relaxation process of prethermalization and thermalization. For $g_1=0.06$ and 0.12, the prethermal dynamics dominates at $\tau\lesssim 5$, and it is followed by a slow relaxation to the thermal equilibrium results. The dynamics in the slow relaxation regime is closely described by the projected dynamics [P-DE, see Eq.~\eqref{eq:O_proj_DE}]. The (small) discrepancy between the actual dynamics and the projected one at late times is quadratic in $g_1$, and its relative magnitude is $\lesssim 10^{-3}$ (see Appendix~\ref{appb} and Ref.~\cite{mallayya2019prethermalization}).

The three scenarios considered in Fig.~\ref{Fig:Pert_dynamics} helps us sharpen the conditions needed to observe a two-step relaxation process. Having in mind that the energy of the perturbed Hamiltonian differs from the energy of the reference Hamiltonian only by an $\mathcal{O}(g)$ correction, one can see that if the perturbation conserves $\hat{N}$ [the case in Fig.~\ref{Fig:Pert_dynamics}(a)] then the ``prethermal'' and the thermal equilibrium states also differ only by an $\mathcal{O}(g)$ correction. Hence, no two-step relaxation process will be seen. 

Breaking a conservation law can make a big difference between the prethermal and thermal results even if the perturbation is small. To understand this, let us focus on the case in which the particle-number conservation is broken. In that case, the thermal equilibrium result is described by $\hat{\rho}^{ }_{\text{GE}}$ in Eq.~\eqref{eq:rho_GE}, which ensures maximal entropy at fixed energy without any constraint on $\hat{N}$, while the prethermal result is given by $\hat{\rho}^{ }_{\text{GE}}$ in Eq.~\eqref{eq:rho_GE_N}, which ensures maximal entropy at fixed energy and fixed particle number. Even if the energy of the perturbed Hamiltonian differs from the energy of the reference Hamiltonian only by an $\mathcal{O}(g)$ correction, the two ensembles are different if $n_I=\text{Tr}[\hat{N}\hat{\rho}^{ }_{I}]/L$ is different from $n^{ }_{{\text{GE}}}(g)=\text{Tr}[\hat{N}\hat{\rho}^{ }_{\text{GE}}]/L$. Namely, observables in the prethermal state will generally be $\mathcal{O}(1)$ different from those in thermal equilibrium if the site occupations have an $\mathcal{O}(1)$ difference. Then, the slow (because $g$ is small) dynamics driven by the perturbation will bring the system from the prethermal equilibrium to the thermal one [Fig.~\ref{Fig:Pert_dynamics}(c)]. However, if the initial state has $n_I=n^{ }_{{\text{GE}}}(g)$, as is the case in Fig.~\ref{Fig:Pert_dynamics}(b) where $n_I=n^{ }_{{\text{GE}}}(g)=0.5$, then the difference between the prethermal and the thermal ensembles is only $\mathcal{O}(g)$, and the dynamics will exhibit a single-step relaxation like the one observed when particle-number conservation is not broken [Fig.~\ref{Fig:Pert_dynamics}(a)].

\begin{figure}[!t]
\includegraphics[width=0.95\linewidth]{ 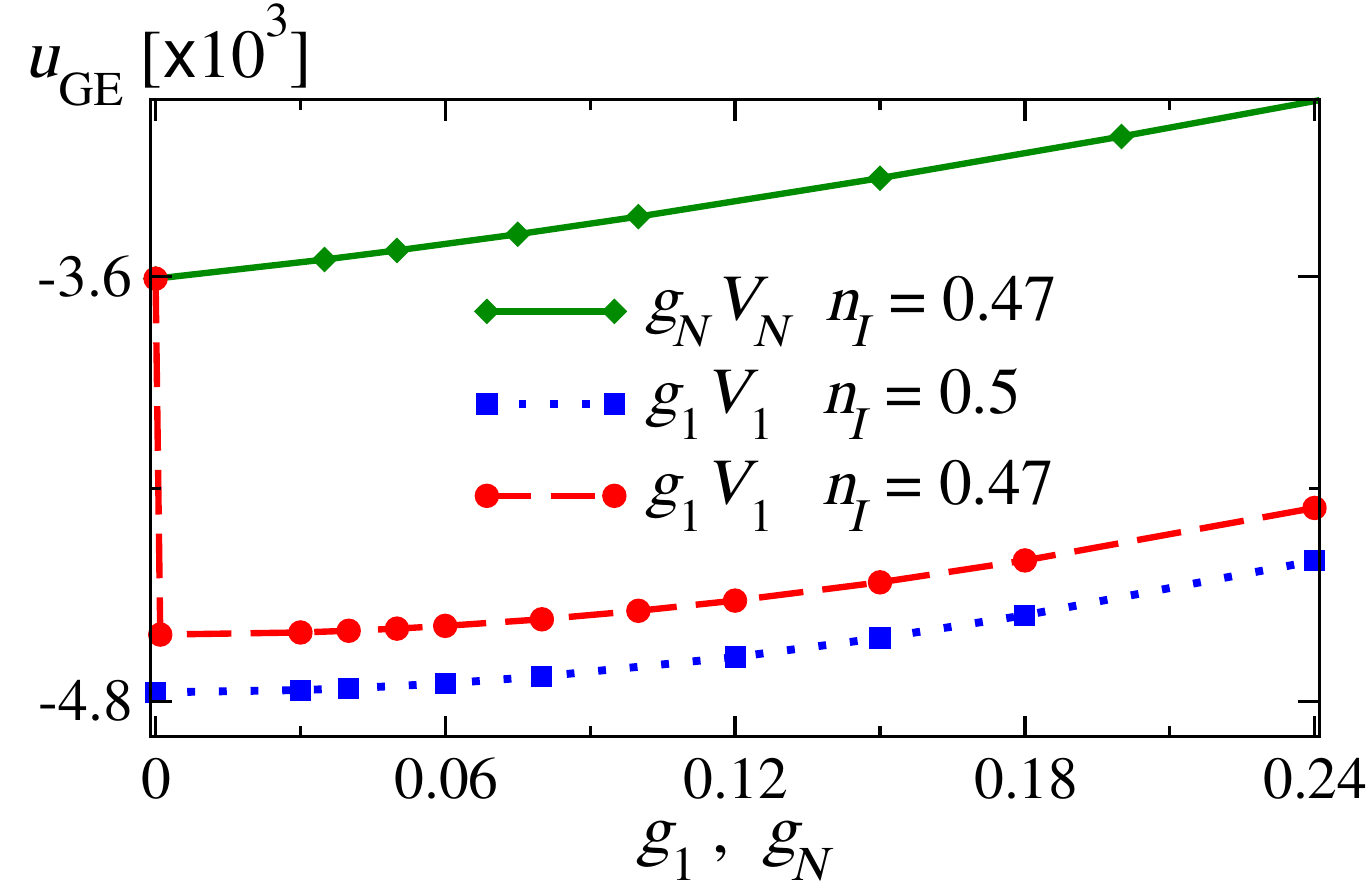}
  \caption{Thermal equilibrium (GE) value of the nearest-neighbor density correlation ($u^{ }_{\text{GE}}$), after evolution under the three scenarios considered in Fig.~\ref{Fig:Pert_dynamics}: $g_N\hat{V}_N$ with $n_I=0.47$ $(\mu_I=1.5)$, $g_1\hat{V}_1$ with $n_I=1/2$ $(\mu_I=0)$, and $g_1\hat{V}_1$ with $n_I=0.47$ ($\mu_I=1.5$). All the results for $u^{ }_{\text{GE}}$ reported in this figure were obtained with the 15$^{\text{th}}$-order NLCE and are converged to machine precision accuracy.}  
\label{Fig:O_GE_vs_g}
\end{figure}

In Fig.~\ref{Fig:O_GE_vs_g}, we show the expectation value of $\hat{U}$ after thermalization, $u^{ }_{\text{GE}} = \text{Tr} [\hat{U} \hat{\rho}^{ }_{\text{GE}}]/L$, as a function of the perturbation strength for the three cases considered in Fig.~\ref{Fig:Pert_dynamics}. Only the red dashed line in Fig.~\ref{Fig:O_GE_vs_g}, corresponding to the GE of the dynamics studied in Fig.~\ref{Fig:Pert_dynamics}(c), shows an $\mathcal{O}(1)$ difference between the prethermal $(g_1=0)$ and the thermal $(g_1>0)$ equilibrium results in Fig.~\ref{Fig:O_GE_vs_g}. This was the dynamics that exhibited two-step relaxation. On the other hand, the two cases that show only a single step relaxation have $\mathcal{O}(g)$ differences between the predictions of the prethermal and the thermal equilibrium ensembles.

While the previous discussion pertained to the thermodynamic limit, the two-step relaxation picture can also be seen in the dynamics of a finite-size system. In the inset in Fig.~\ref{Fig:Pert_dynamics}(c), we show the relaxation of $u(\tau)$ under $\hat{H}_{g_1}$ (also when the initial state is away from half filling), in a finite chain with $19$ sites and periodic boundary conditions. Those results were obtained using exact diagonalization (ED). $u(\tau)$ for the $19$-site chain looks qualitatively similar to the thermodynamic limit result, with an early fast prethermal dynamics followed by a slow relaxation to the DE of the finite-size system. The slow relaxation regime follows the projected dynamics given by the P-DE of the finite chain. That said, as we discuss in Sec.~\ref{sec:finsize}, finite-size effects can lead to misleading conclusions about the dependence of the relaxation rates on the strength of the perturbation so care should be taken when studying finite-size systems.

\KM{Another point to be highlighted about the results reported in Fig.~\ref{Fig:Pert_dynamics} is that they make apparent the need to sharpen analyses reported in recent papers that explored prethermalization and thermalization in the context of random matrix theory and typicality~\cite{reimann2019typicality, dabelow2020modification, dabelow2020relaxation, *dabelow2021typical, richter_20, heitmann2021nontrivial}. In those works prethermalization was argued to be generic in systems with perturbed dynamics, with no explicit conditions on the nature of the perturbations.}

\subsection{Projected Dynamics}\label{sec_proj_dy}

\begin{figure}[!t]
\includegraphics[width=0.985\linewidth]{ 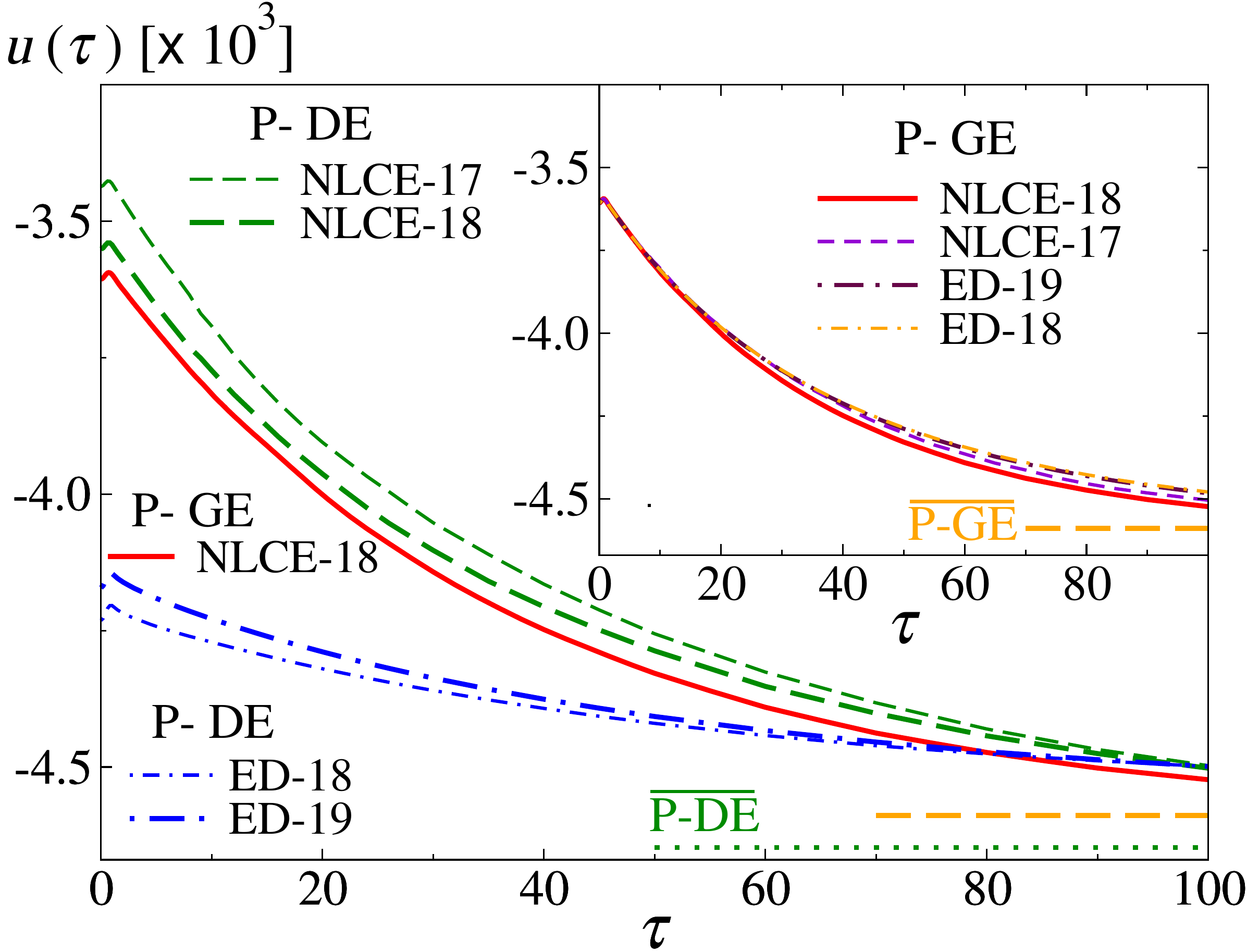}
  \caption{Projected dynamics of $u(\tau)$ following the quench $\hat{H}_I\rightarrow\hat{H}_{g_1}$, with $g_1=0.06$. The initial state has $\mu_I=1.5$ ($n_I=0.47$). The main panel shows the projected DE (P-DE) results evaluated with NLCE to 17$^{\text{th}}$-order (NLCE-17) and 18$^{\text{th}}$-order (NLCE-18), and ED in chains with $18$ sites (ED-18) and 19 sites (ED-19) and periodic boundary conditions. The equilibrium value of the projected DE $(\overline{\text{P-DE}})$ [Eq.~\eqref{eq:Obar_PDE}] is evaluated with NLCE-18 and shown as a horizontal dotted line. The inset shows the projected GE (P-GE) results obtained using NLCE-17 and NLCE-18 (the P-GE results for NLCE-18 are also shown in main panel), as well as ED-18 and ED-19. The thermal equilibrium result of the projected dynamics $(\overline{\text{P-GE}})$ [Eq.\eqref{eq:O_PGE_bar}] are shown in the main panel and the inset as a horizontal dashed line (computed within machine precision using the 15$^{\text{th}}$-order NLCE).}
\label{Fig:Proj_Dynamics}
\end{figure}

Having identified two essential conditions to observe a two-step relaxation dynamics (under perturbation $g_1\hat{V_1}$ and an initial state with $n_I\ne1/2$), we focus next on the slow thermalization regime described by the projected dynamics of $\hat{U}$. In the thermodynamic limit, the projected dynamics can be described equivalently with a P-DE [Eq.~\eqref{proj_DE}] or a P-GE [Eq.~\eqref{eq:rho_Proj_GE}]. The final equilibrium value of the projected dynamics can be evaluated equivalently with $\overline{\text{P-DE}}$ defined in Eq.~\eqref{eq:Obar_PDE} or its thermal counterpart $\overline{\text{P-GE}}$ [Eq.~\eqref{eq:O_PGE_bar}]. 
\KM{In Ref.~\cite{mallayya2019prethermalization}, the projected dynamics were computed only using the diagonal ensemble (P-DE). In Fig.~\ref{Fig:Proj_Dynamics}, we show results for the projected dynamics in both the DE and the GE} evaluated within the last two orders (17 and 18) of the NLCE, and in the two largest chains (18 and 19 sites) with periodic boundary conditions solved using ED. 

Let us first focus on the inset in Fig.~\ref{Fig:Proj_Dynamics}, which shows results for P-GE evaluated using NLCE and ED. To evaluate P-GE at the $l^{\text{th}}$-order NLCE ($L$ sites with ED), the dynamics of the reference energy $e_0(\tau)=\text{Tr}[\hat \rho(\tau)\hat H_0]/L$ and particle number $n(\tau)=\text{Tr}[\hat \rho(\tau)\hat N]/L$ are evaluated using the $l^{\text{th}}$-order NLCE ($L$ sites with ED). From $e_0(\tau)$ and $n(\tau)$, the temperature $T(\tau)$ and chemical potential $\mu(\tau)$ that define $\hat{\rho}_{\text{P-GE}}(\tau)$ are determined by numerically solving Eqs.~\eqref{eq:proj_finalT} and~\eqref{eq:proj_finalmu} at each $\tau$. The two equations are solved to an accuracy of relative errors $\lesssim 10^{-11}$. With $\hat{\rho}^{ }_{\text{P-GE}}(\tau)$ thus obtained, the P-GE of $\hat{U}$ is evaluated to machine precision with NLCE. With all other calculations converged to machine precision, the only source of error in the P-GE calculations stems from the convergence errors (finite-size errors) in obtaining $e_0(\tau)$ and $n(\tau)$ in the dynamics with the $l^{\text{th}}$-order NLCE ($L$ sites with ED). The inset in Fig.~\ref{Fig:Proj_Dynamics} shows that the errors in the P-GE dynamics of $\hat{U}$ are apparent only at long times ($\tau \gtrsim20$). At earlier times, the relative error between the 18$^{\text{th}}$-order NLCE and the $L=19$ sites' ED calculation is less than $10^{-7}$ for $\tau <2$, $10^{-5}$ for $\tau<3$ and $10^{-3}$ for $\tau\lesssim 10$. Thus the P-GE at short times is accurately estimated both using NLCE and ED.

In the main panel in Fig.~\ref{Fig:Proj_Dynamics}, we show the projected dynamics of $\hat{U}$ given by the P-DE, evaluated within the 17$^{\text{th}}$ and 18$^{\text{th}}$-order NLCE, $L=18$ and 19 sites using ED, as well as the P-GE result from the 18$^{\text{th}}$-order NLCE (the same curve shown in the inset). The results in the main panel in Fig.~\ref{Fig:Proj_Dynamics} show that the P-DE results from the ED calculations exhibit much larger deviations from the P-GE results at short times than the P-DE results from the NLCE calculations (we remind the reader that the P-GE results at short times exhibit negligible convergence errors, as shown in the inset). With increasing $L$, the P-DE results approach the ones of the P-GE, albeit slowly. With increasing the order of the NLCE, the already closer NLCE results for the P-DE at short times approach rapidly the ones of the P-GE. 

The thermal equilibrium result of the projected dynamics  $\overline{\text{P-GE}}$ (expected to be reached at long times), is given by Eq.~\eqref{eq:O_PGE_bar} where $\bar{T}$ and $\bar{\mu}$ are obtained numerically using Eqs.~\eqref{eq:proj_finalT_bar} and~\eqref{eq:proj_finalmu_bar}. For the expectation value of $\hat{U}$ in $\overline{\text{P-GE}}$, $\bar{u}^{ }_{\text{P-GE}}$, no step in the calculation involves dynamics and all the density matrices are Gibbs ensembles, which means that one can use the NLCE to calculate $\bar{u}^{ }_{{\text{P-GE}}}$ exactly to machine precision. The result is shown in the main panel and the inset in Fig.~\ref{Fig:Proj_Dynamics} as a dashed horizontal line marked $\overline{\text{P-GE}}$. The equilibrium value of the P-DE predicted by $\overline{\text{P-DE}}$ [Eq.~\eqref{eq:Obar_PDE}] for $\hat{U}$, $\bar{u}^{ }_{\text{P-DE}}$ evaluated with NLCE is shown in the main panel as a dotted horizontal line. The discrepancy between $\bar{u}^{ }_{\text{P-DE}}$ and $\bar{u}^{ }_{\text{P-GE}}$ is due to the convergence errors of the former. Like other DE predictions, these errors decrease with increasing the order of the NLCE (see Appendix~\ref{appa}).

From the above analysis, we conclude that the slow relaxation regime of observables in the thermodynamic limit is most accurately described by the P-GE at short times, and by $\overline{\text{P-GE}}$ at $\tau\rightarrow\infty$. The P-DE results evaluated with NLCE are close and approach rapidly the P-GE results at short times. On the other hand, the P-DE results obtained using ED in a periodic chain with $L$ sites exhibit large deviations from the P-GE results at short times, and approach the P-GE results slowly with increasing $L$.

\begin{figure*}[!t]
\includegraphics[width=0.98\linewidth]{ 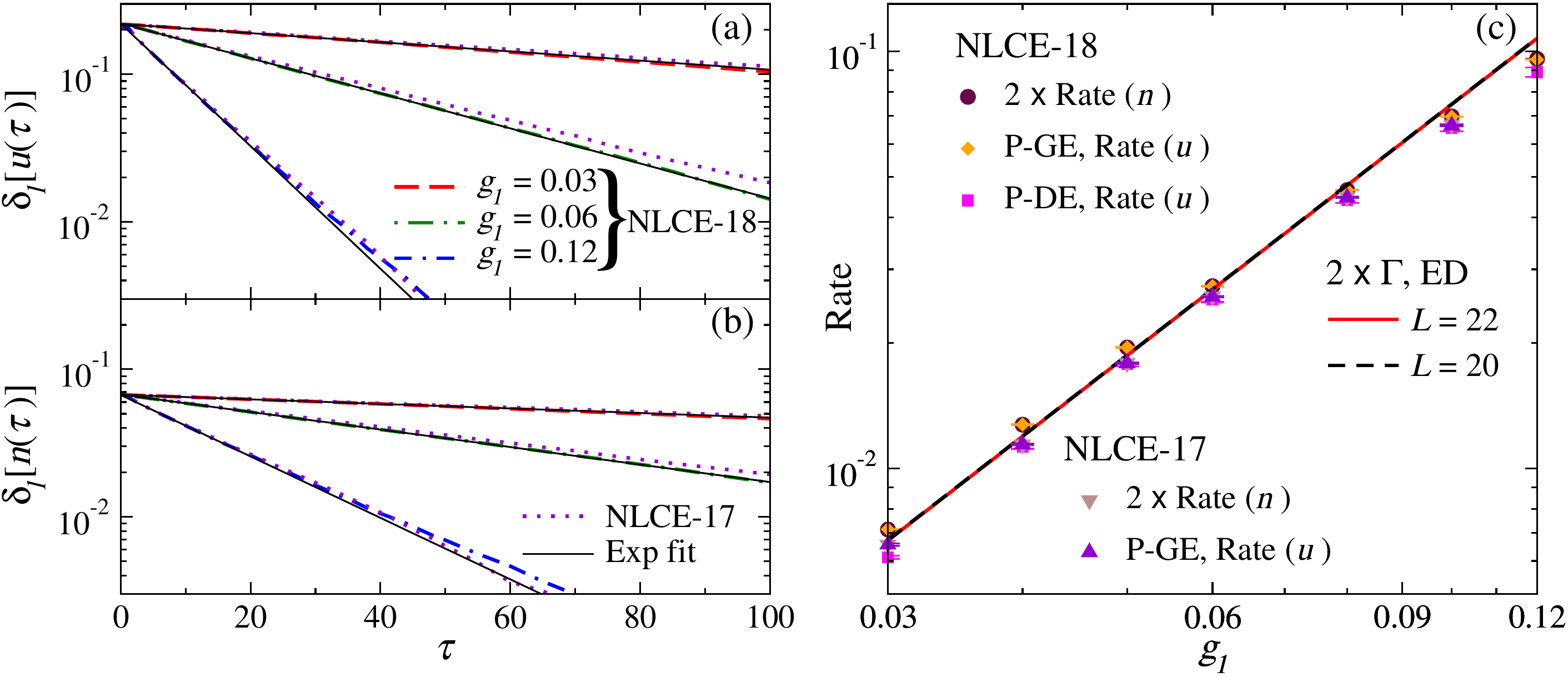}
  \caption{Slow relaxation dynamics of $u(\tau)$ and $n(\tau)$ as described by projected-dynamics calculations following a quench $\hat{H}_I\rightarrow\hat{H}_{g_1}$ in the thermodynamic limit. The initial state is taken to have $\mu_I=1.5$ and $T_I=10$. In (a) we show P-GE results for $u(\tau)$ and in (b) for $n(\tau)$. In both panels exponential relaxation is captured by the distance to thermalization $\delta_l[u(\tau)]$ [Eq.~\eqref{eq:delta_l_u}] and $\delta_l[n(\tau)]$ [Eq.~\eqref{eq:delta_l_n}], respectively. The results in (a) and (b) are evaluated within the $18^{\text{th}}$ (NLCE-18) and $l=17^{\text{th}}$ (NLCE-17) orders of the NLCE, for three perturbation strengths, $g_1=0.03,\,0.06,$ and 0.12. The solid lines are exponential fits to the NLCE-18 results in the time interval $3\le\tau\le 16$. (c) The relaxation rates obtained from exponential fits to $\delta_l[u(\tau)]$ [Rate($u$)] and $\delta_l[n(\tau)]$ [Rate($n$)] with NLCE-18 and NLCE-17 in the P-GE, as well as the Rate($u$) obtained from fits to the P-DE with NLCE-18, for various values of $g_1$. All exponential fits were done for $3\le\tau\le 16$. The lines give the rate, $\Gamma$, predicted by the FGR formula [Eq.~\eqref{eq:FGR_rate_N}] evaluated using ED for a periodic chain with $L=22$ (solid line) and $L=20$ sites (dashed line). Rate($n$) and $\Gamma$ are multiplied by a factor 2 to be compared with Rate($u$), see text.}
\label{Fig:Rate_NLCE}
\end{figure*}

\subsection{Relaxation rates in the thermodynamic limit}\label{sec_rate_NLCE} 

Here we analyze the slow thermalization of $\hat{U}$ and $\hat{N}$ for the quench $\hat{H}_I\rightarrow\hat{H}_{g_1}$ in which the initial state has $\mu_I=1.5$ ($n_I=0.47$). This slow relaxation regime is described by the projected dynamics. We showed in the previous section that, in the thermodynamic limit, the projected dynamics and its long-time equilibrium are described most accurately by the P-GE [Eq.~\eqref{eq:O_PGE}] and $\overline{\text{P-GE}}$ [Eq.~\eqref{eq:O_PGE_bar}], respectively, evaluated with NLCE. 

In Fig.~\ref{Fig:Rate_NLCE}(a), we show that the relaxation dynamics of $\hat{U}$ predicted by the P-GE is well described by an exponential for $g_1\lesssim 0.12$. We show results for the normalized distance to thermalization:
\begin{equation}
\delta_l[u(\tau)]=\left|\dfrac{u^{l}_{{\text{P-GE}}}(\tau)-\bar{u}^{ }_{{\text{P-GE}}}}{\bar{u}^{ }_{{\text{P-GE}}}}\right|, \label{eq:delta_l_u}
\end{equation}
where $u^{l}_{{\text{P-GE}}}(\tau)$ is the prediction of the P-GE at time $\tau$ evaluated at the $l^{\text{th}}$-order NLCE, and $\bar{u}^{ }_{{\text{P-GE}}}$ is the thermal equilibrium value of the projected dynamics evaluated to machine precision.

Similarly, the broken conserved quantity $\hat{N}$ relaxes exponentially as shown in Fig.~\ref{Fig:Rate_NLCE}(b), with its distance to thermalization $\delta_l[n(\tau)]$ given by
\begin{equation}
\delta_l[n(\tau)]=\left|\dfrac{n^{ }_{l}(\tau)-n^{ }_{{\text{GE}}}}{n^{ }_{{\text{GE}}}}\right|, \label{eq:delta_l_n}
\end{equation}
where $n^{ }_{l}(\tau)$ is the particle number at $\tau$ evaluated at the $l^{\text{th}}$-order NLCE and $n^{ }_{{\text{GE}}}=1/2$ is the equilibrium site occupation. For both $\delta_l[u(\tau)]$ and $\delta_l[n(\tau)]$, the last two orders of the NLCE are well converged at short times and fit well to an exponential (from which we extract the relaxation rate). To fit the exponential, we select a time window $3\le\tau\le16$ that excludes long times at which the convergence is not as good, and early times at which transient prethermal dynamics is present. The rates thus extracted for $0.03\le g_1\le0.12$ are shown in Fig.~\ref{Fig:Rate_NLCE}(c).

A well defined, time-independent, relaxation rate of $\hat{N}$ can be obtained using the drift $dn/d\tau$ given by the FGR formula [Eq.~\eqref{fermi_rate}] when $n(\tau)$ is sufficiently close to $n^{ }_{{\text{GE}}}$. This eventually occurs at long times for any $n_I\ne n^{ }_{{\text{GE}}}$. In the case of nonintegrable $\hat{H}_0$ and the perturbation $g_1\hat{V}_1$ [Eq.~\eqref{H1}], we can simplify Eq.~\eqref{fermi_rate} using the P-GE, and obtain the $\tau$ independent relaxation rate (see Appendix~\ref{appc}). The FGR equation involves two separate contributions from the perturbation: $\hat{V}_1=\hat{\mathcal{V}}_{\eta=1}+ \hat{\mathcal{V}}_{\eta=2}$, where $\hat{\mathcal{V}}_{\eta}$ connects the eigenkets of $\hat{H}_0$ that differ by $\eta$ particles. In a block diagonalized symmetry sector $s$ of both $\hat{H}_{0}$ and $\hat{N}$, let $F^s_{N,\eta}(E_0)$ be the coarse grained value of the squared matrix elements of $\hat{\mathcal{V}}_{\eta}$ given by
\begin{equation}
F^s_{N,\eta}(E_0)=\text{Avg}_{\Delta E}\left(|\bra{E^{N+\eta}_j}\hat{\mathcal{V}}_{\eta}\ket{E^{N}_i}|^2\right),\label{eq:Vij_sq}
\end{equation}
where $\ket{E^{N}_i}$ and $\ket{E^{N+\eta}_j}$ are the eigenkets of $\hat{H}_0$ with energies within a small window $(E_0-\Delta E/2, E_0+\Delta E/2)$, and particle number $N$ and $N+\eta$, respectively. The FGR rate $\Gamma$ for this system, when $L\gg 1$, is given by $\Gamma=\Gamma_{(\eta=1)}+\Gamma_{(\eta=2)}$ with
\begin{eqnarray}
\Gamma_{\eta}=&&\dfrac{2\pi\eta^2g_1^2}{\text{Tr}[\hat{N}^2 e^{-\bar{\beta}\hat{H}_0}]}\sum_{s}\sum_{N=0}^{L-\eta}\int dE_0\  e^{-\bar{\beta} E_0}\label{eq:FGR_rate_N}\nonumber\\ &&\times F^s_{N,\eta}(E_0)D^{s}_{N+\eta}(E_0)D^{s}_{N}(E_0)\label{eq:fgrtl}
\end{eqnarray}
where $\bar{\beta}^{-1}=\bar{T}$ is the temperature of the $\overline{\text{P-GE}}$ [Eq.~\eqref{eq:proj_finalT_bar}] in the limit of $g_1\rightarrow 0$, $D^s_N(E_0)$ is the density of states of $\hat{H}_0$ at energy $E_0$, particle number $N$, in sector $s$, and $F^s_{N,\eta}(E_0)$ is defined in Eq.~\eqref{eq:Vij_sq}. For $\Gamma$ in Eq.~\eqref{eq:FGR_rate_N} to be well defined in the thermodynamic limit $F^s_{N,\eta}(E_0)$ needs to be $\propto [D^{s}_{N}(E_0)]^{-1}$, where we used that $D^{s}_{N+\eta}(E_0)\simeq D^{s}_{N}(E_0)$ in large systems. This is expected to be the case both in nonintegrable and in integrable-interacting systems for local operators that connect different sectors of a Hamiltonian~\cite{leblondrigol2020}. We note that Eq.~\eqref{eq:fgrtl} has similarities with the heating rate formula derived in Ref.~\cite{mallayya2019heating} for periodically driven perturbations. 

We calculate $\Gamma$ evaluating Eq.~\eqref{eq:FGR_rate_N} numerically for a finite system with periodic boundary conditions using ED (see Appendix~\ref{appd}). Since this calculation does not involve dynamics and requires only the diagonalization of $\hat{H}_0$, which conserves $\hat{N}$, we are able to compute the rates using a larger periodic chain (with $22$ sites). In Fig.~\ref{Fig:Rate_NLCE}(c), we show that the rates of $\hat{N}$ [Rate($n$)] extracted from exponential fits like the ones shown in Fig.~\ref{Fig:Rate_NLCE}(b) are in excellent agreement with $\Gamma$ evaluated using Eq.~\eqref{eq:FGR_rate_N}. 

On the other hand, the rates of $\hat{U}$ [Rate($u$)] estimated from fits of the P-GE results [as in Fig.~\ref{Fig:Rate_NLCE}(a)] are twice as large. This factor of 2 can be understood by noticing that in the P-GE dictated by the evolution of $n(\tau)$, the particle-hole symmetry of $\hat{U}$ implies that $\left( u^{l}_{{\text{P-GE}}} (\tau)-\bar{u}^{ }_{{\text{P-GE}}} \right) \propto \left(n(\tau)-0.5\right)^2 + \mathcal{O} \left( \left|n(\tau)-0.5\right|^4 \right)$\KM{, when $n(\tau)$ is close to 0.5 [so that an expansion in powers of $n(\tau)-0.5$ is meaningful]}. Thus $\delta_l[n(\tau)]\propto \exp\left(-\Gamma \tau\right)$ results in $\delta_l[u(\tau)]\propto \exp \left(-2\Gamma \tau\right)$. The Rate($u$) and $2\times$Rate($n$) estimated from exponential fits to the P-GE dynamics evaluated using 17 and 18 orders of the NLCE, and $2\Gamma$ evaluated with FGR in chains with $L=20$ and $L=22$ sites show excellent agreement with each other, with small discrepancies developing at larger $g_1\sim 0.12$ (where FGR becomes less accurate). 

In Fig.~\ref{Fig:Rate_NLCE}(c), we also show the thermalization rates for $\hat{U}$ obtained using fits to the P-DE dynamics, $u^{ }_{{\text{P-DE}}}(\tau)$, evaluated at the $18^{\text{th}}$-order NLCE. Those rates are computed using exponential fits to the equivalent of Eq.~\eqref{eq:delta_l_u} in the P-DE, involving $u^{ }_{{\text{P-DE}}}(\tau)$ and the equilibrium value $\bar{u}^{ }_{{\text{P-DE}}}$ [Eq.~\eqref{eq:Obar_PDE}]. The P-DE rates are in good agreement with those obtained in the other calculations, as expected given the results in Fig.~\ref{Fig:Proj_Dynamics}.

\section{Relaxation rates in finite systems}\label{sec:finsize} 

In this section, we discuss what happens when one uses finite-system calculations to study relaxation rates during dynamics at even smaller values of the perturbation strength than the ones considered in Fig.~\ref{Fig:Rate_NLCE}. Since the relaxation dynamics is very slow in that regime, one needs long times to be able to fit exponentials and extract relaxation rates. This is something that can always be done using exact diagonalization in finite systems. On the other hand, because of the lack of convergence of NLCE calculations at long times, this is a regime that cannot be studied using NLCEs,

\begin{figure*}[!t]
\includegraphics[width=0.99\linewidth]{ 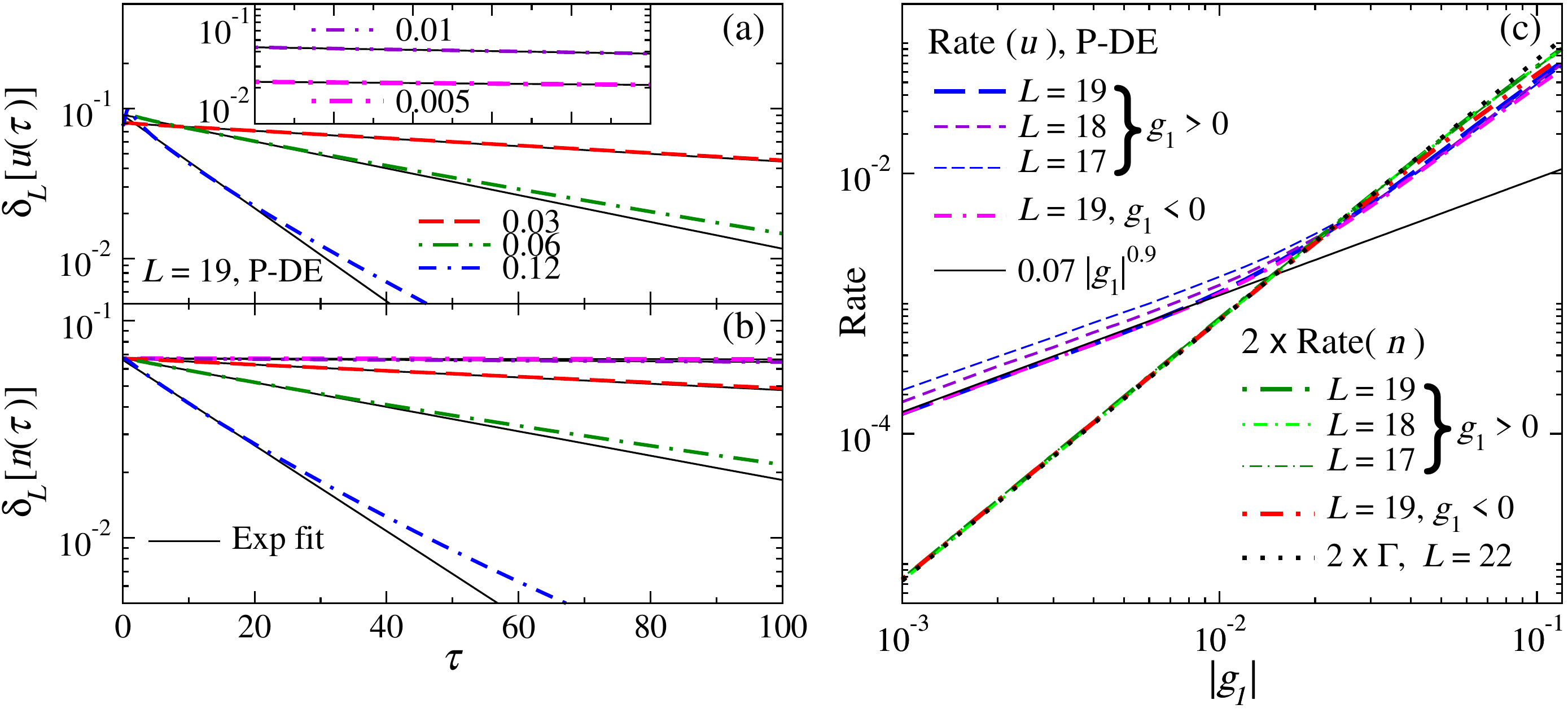}
\caption{The projected dynamics of $u(\tau)$ and $n(\tau)$ as described by the P-DE in finite chains with $L$ sites and periodic boundaries. The exponential relaxation of (a) $u(\tau)$ and (b) $n(\tau)$ is captured by the distance to equilibrium, $\delta_L[u(\tau)]$ [Eq.~\ref{eq:delta_L_u}] and $\delta_L[n(\tau)]$  [Eq.~\ref{eq:delta_L_n}] respectively. The results in (a) and (b) are obtained in chains with $L=19$ sites, for the perturbation strengths $g_1=0.005,\, 0.01,\, 0.03,\,0.06,$ and 0.12. For clarity, the results for $\delta_L[u(\tau)]$ when $g_1=0.005$ and 0.01 are shown as an inset in (a). The solid lines are exponential fits in the time interval $3\le\tau\le 20$. (c) The relaxation rates obtained from exponential fits to $\delta_L[u(\tau)]$ [Rate($u$)] and $\delta_L[n(\tau)]$ [Rate($n$)] for $L=17,\,18,$ and 19 sites, and various \KM{positive and negative} values of $g_1$ \KM{($0.001\le |g_1|\le 0.12$)}. All exponential fits were done for $3\le\tau\le 20$. The solid line shows the result of a power-law fit \KM{$\alpha |g_1|^{\beta}$ to $\delta_{19}[u(\tau)]$ for $10^{-3}\le |g_1|\le10^{-2}$}. The dotted line shows the FGR rate $\Gamma$ [Eq.~\eqref{eq:FGR_rate_N}] evaluated using ED for $L=22$ sites. Rate($n$) and $\Gamma$ are multiplied by a factor 2 to be compared with Rate($u$), see text.}\label{Fig:Rate_ED}
\end{figure*}

We study the P-DE dynamics of $\hat{U}$ and $\hat{N}$ for the same quench discussed in Sec.~\ref{sec_rate_NLCE}, but explore much smaller perturbations $g_1\in(0.001, 0.12)$. As shown in the inset in Fig.~\ref{Fig:Pert_dynamics}(c), the P-DE dynamics in finite systems closely follows the exact dynamics, but it can be significantly different from the P-GE dynamics (see Fig.~\ref{Fig:Proj_Dynamics}). The differences are expected to increase in finite systems for smaller perturbations, because the assumption of thermalization at each time during the slow relaxation (to a new thermal equilibrium determined by the instantaneous site occupation) is less justified. Hence, in this section, we do not assume thermalization and use the P-DE [Eq.~\eqref{eq:O_proj_DE}] to describe dynamics and $\overline{\text{P-DE}}$ [Eq.~\eqref{eq:Obar_PDE}] to describe the final equilibrium state of the P-DE. 

In Fig.~\ref{Fig:Rate_ED}(a), we show how the expectation value of $\hat{U}$ equilibrates in finite systems for different values of $g$. There we plot the distance to equilibrium  
\begin{equation}
\delta_L[u(\tau)]=\left|\dfrac{u^{L}_{{\text{P-DE}}}(\tau)-\bar{u}^{L }_{{\text{P-DE}}}}{\bar{u}^{L}_{{\text{P-DE}}}}\right|, \label{eq:delta_L_u}
\end{equation}
where $u^{L}_{{\text{P-DE}}}(\tau)$ is the expectation value in the P-DE at time $\tau$, and $\bar{u}^{L} _{{\text{P-DE}}}$ is the equilibrated result predicted by $\overline{\text{P-DE}}$, for a chain with $L$ sites. Similarly, in Fig.~\ref{Fig:Rate_ED}(b), we show the equilibration dynamics of $\hat{N}$ as characterized by the distance to equilibrium
\begin{equation}
\delta_L[n(\tau)]=\left|\dfrac{n^{ }_{L}(\tau)-n^{ }_{{\text{DE}}}}{n^{ }_{{\text{DE}}}}\right|, \label{eq:delta_L_n}
\end{equation}
where $n^{ }_{L}(\tau)$ is the particle number (per site) at $\tau$, and the equilibrium site occupation $n^{ }_{{\text{DE}}}=1/2$ (because of the particle-hole symmetry of $\hat{H}_{g_1}$). 

Figures~\ref{Fig:Rate_ED}(a) and~\ref{Fig:Rate_ED}(b) show that both $\delta_L[u(\tau)]$ and $\delta_L[n(\tau)]$ exhibit exponential regimes, for a chain with $L=19$ sites, as those identified in Figs.~\ref{Fig:Rate_NLCE}(a) and~\ref{Fig:Rate_NLCE}(b) in the NLCE calculations for the thermodynamic limit. A fit to $\delta_L[u(\tau)]$ and $\delta_L[n(\tau)]$ in the time window $3\le\tau\le 20$ agrees well with an exponential. The relaxation rates thus obtained from exponential fits for $\hat{U}$ [Rate($u$)] and $\hat{N}$ [Rate($n$)] for various  \KM{positive and negative} values of $g_1$  and chains with three different sizes $L$, along with the FGR rate $\Gamma$ from Fig.~\ref{Fig:Rate_NLCE} for $L=22$, are shown in Fig.~\ref{Fig:Rate_ED}(c).      

In Fig.~\ref{Fig:Rate_ED}(c), the results for Rate($u$) exhibit a surprising (and potentially misleading) finite size effect. The Rate($u$) for \KM{$|g_1|<0.01$} deviates significantly away from the expected FGR prediction of $2\Gamma$, and scales almost linearly with \KM{$|g_1|$}. The latter behavior is different from the signature $(g_1)^2$ scaling of the FGR prediction. Increasing $L$ brings the Rate($u$) slowly towards $2\Gamma$ by pushing the spurious scaling regime to smaller values of \KM{$|g_1|$}. Currently, we do not understand why such a linear scaling regime emerges in the Rate($u$) at small $|g_1|$ in finite-system calculations. A recent study exploring how integrability and Anderson localization are broken in finite systems showed that a highly nontrivial regime precedes the onset of quantum chaos~\cite{leblond2020}. We expect similar nontrivial finite-size effects when breaking only one conservation law. Another point to be noted about the results for the Rate($u$) in Fig.~\ref{Fig:Rate_ED}(c) is that, as expected because of finite-size effects, the deviations from the FGR predictions are larger than those seen in the NLCE calculations in Fig.~\ref{Fig:Rate_NLCE} for $0.03\le g_1\le0.12$.

Contrary to the results for the Rate($u$), the results for the Rate($n$) in Fig.~\ref{Fig:Rate_ED}(c) do not show significant finite-size effects, and agree well with the FGR rate $\Gamma$ for all values of $g_1$ and $L$ considered. This can be understood by noticing that $n^{ }_{L}(\tau)$, being the conserved quantity of $\hat{H}_0$, has the same value in the P-DE (by definition) and the P-GE (by construction) as for the actual dynamics under $\hat{H}_{g_1}$. Since the actual dynamics at short times has small finite-size effects (because of locality at short times the system does not ``know'' its extent), then $n^{ }_{L}(\tau)$ at short times agrees well with its value in the thermodynamic limit. At long times, both in finite systems and in the thermodynamic limit, $n^{ }_{{\text{DE}}}=1/2$ for all values of $g_1$ because this is set by the particle-hole symmetry. Hence, the thermodynamic limit behavior of $\delta_L[n(\tau)]$ at short times is properly captured by the exact diagonalization calculations and so are the values of Rate($n$) obtained from the (short-)time evolution fits. 

\section{Summary}\label{sec:summary}

We used numerical simulations to carry out an in-dept analysis on the framework of prethermalization and thermalization introduced in Ref.~\cite{mallayya2019prethermalization}. We considered far-from-equilibrium initial states evolving under  Hamiltonians of the form $\hat{H}_0+g\hat{V}$, where $\hat{H}_0$ is nonintegrable and has one extensive conserved quantity (the total number of particles). We studied the dynamics of local observables in the thermodynamic limit using a NLCE, and in finite chains with periodic boundary conditions using ED.    

By exploring three distinct scenarios with different perturbations and initial states, we showed that in order for a two-step relaxation process to occur in the dynamics of local observables (other than the conserved quantity), not only does the perturbation $g\hat{V}$ have to break the extensive conserved quantity of $\hat{H}_0$, but also the initial-state expectation value of the conserved quantity per site needs to be $\mathcal{O}(1)$ different from the final equilibrium one. Observables (other than the conserved quantity) then evolve with a fast prethermal dynamics at short times followed by a slow relaxation to thermal equilibrium at long times. The slow relaxation regime is characterized by intermediate equilibrium states of $\hat{H}_0$, which in the thermodynamic limit can be equivalently described using the projected DE (P-DE) or the projected GE (P-GE).

We argued that the thermodynamic limit results for the slow thermalization regime are most accurately described numerically using NLCE calculations for the P-GE. Using such calculations we showed that the slow thermalization \KM{regime} is exponential, with a rate that can be accurately predicted using Fermi's golden rule [Eq.~\eqref{eq:FGR_rate_N}]. We also showed that the NLCE results for the rates obtained using the P-DE are in good agreement with the P-GE and Fermi's golden rule ones. 

On the other hand, in a finite system and for quantities that are not conserved in the reference dynamics, such as the expectation value of the nearest-neighbor density correlator, we showed that the P-DE calculations exhibit large finite-size effects. Strikingly, for very small perturbations, finite-size effects result in rates that are linear \KM{in the absolute value} of the perturbation strength, at odds with Fermi's golden rule prediction. Increasing the system size pushes this linear scaling regime to smaller values of the perturbation so that it disappears in the thermodynamic limit. 

\begin{acknowledgements}
This work was supported by the U.S. Office of Naval Research, Grant No.~N00014-14-1-0540 (K.M.), and by the National Science Foundation, Grant No.~2012145 (M.R.). The computations were carried out at the Institute for CyberScience at Penn State.
\end{acknowledgements}

\appendix

\section{Convergence of the DE and GE calculations with NLCE}\label{appa}

Here we discuss the convergence of the NLCE calculations for the expectation value of $\hat{U}$ after thermalization, as described by the DE [Eq.~\eqref{O_DE}] and GE [Eq.~\eqref{O_GE}], as well as the corresponding predictions of the projected dynamics of $\hat{U}$ described by the $\overline{\text{P-DE}}$ [Eq.~\eqref{eq:Obar_PDE}] and $\overline{\text{P-GE}}$ [Eq.~\eqref{eq:O_PGE_bar}]. 

\begin{figure}[!b]
\includegraphics[width=0.985\linewidth]{ 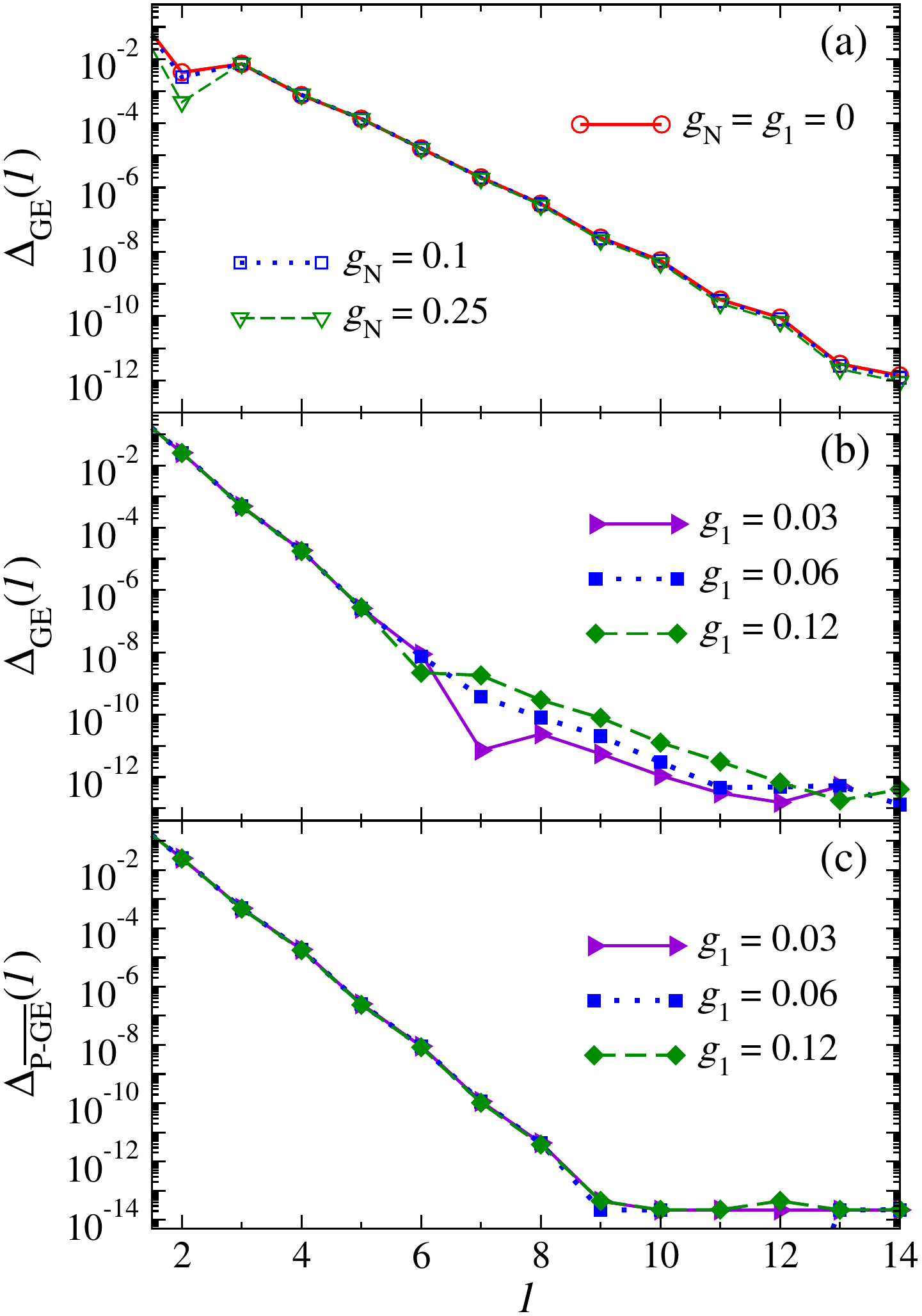}
  \caption{Convergence of the NLCE calculations for the expectation value of $\hat{U}$ in the GE following the quenches: (a) $\hat{H}_I\rightarrow \hat{H}_{g_N}$ and (b) $\hat{H}_I\rightarrow \hat{H}_{g_1}$. (c) Convergence of the $\overline{\text{P-GE}}$ predictions for the quench $\hat{H}_I\rightarrow \hat{H}_{g_1}$ [same quench as in (b)]. The initial state has $T_I=10$ and $\mu_I=1.5$ ($n_I=0.47$). The convergence error at the $l^{\text{th}}$-order is quantified by the relative deviation from the highest order $(l=15)$ NLCE estimate, given by $\Delta_{\text{GE}}(l)$ for the GE and $\Delta_{\overline{\text{P-GE}}}(l)$ for the $\overline{\text{P-GE}}$; see Eq.~\eqref{eq:delta_GE_nlce}.}  
\label{Fig:GE_conv}
\end{figure}

In Fig.~\ref{Fig:GE_conv}, we show results that exemplify how the NLCE calculations converge, with increasing the NLCE order $l$, for the expectation value of $\hat{U}$ in the GE $(u^{l}_{{\text{GE}}})$ and in  the $\overline{\text{P-GE}}$ $(\bar{u}^{l}_{{\text{P-GE}}})$. The errors are quantified by the relative differences
\begin{equation}
\Delta_{\text{GE}}(l)=\left|\dfrac{u^{l}_{{\text{GE}}}-u^{15}_{{\text{GE}}}}{u^{15}_{{\text{GE}}}}\right|,\  \Delta_{\overline{\text{P-GE}}}(l)=\left|\dfrac{\bar{u}^{l}_{{\text{P-GE}}}-\bar{u}^{15}_{{\text{P-GE}}}}{\bar{u}^{15}_{{\text{P-GE}}}}\right| ,\label{eq:delta_GE_nlce}
\end{equation}
where $l=15$ is the highest order computed for the GE and for the $\overline{\text{P-GE}}$. Figures~\ref{Fig:GE_conv}(a) and~\ref{Fig:GE_conv}(b) show $\Delta_{\text{GE}}(l)$ for the 
quenches $\hat{H}_I\rightarrow\hat{H}_{g_N}$ and  $\hat{H}_I\rightarrow\hat{H}_{g_1}$, respectively, while Fig.~\ref{Fig:GE_conv}(c) shows $\Delta_{\overline{\text{P-GE}}}(l)$ for the latter quench. In all the cases, the errors decrease exponentially with increasing $l$, and the relative errors between the $14^{\text{th}}$- and the $15^{\text{th}}$-order NLCE are $\lesssim 10^{-12}$. This is generic for NLCE calculations within the grand canonical ensemble at finite (not too low) temperature~\cite{iyer2015optimization}. The temperature (and chemical potential when $\hat{N}$ is conserved) of the GE and $\overline{\text{P-GE}}$ are set by the initial state and they are computed (within a grand canonical ensemble) so that the relative errors are $\lesssim 10^{-12}$. Given those errors, we can consider the GE and $\overline{\text{P-GE}}$ results exact to machine precision. 

\begin{figure}[!t]
\includegraphics[width=0.985\linewidth]{ 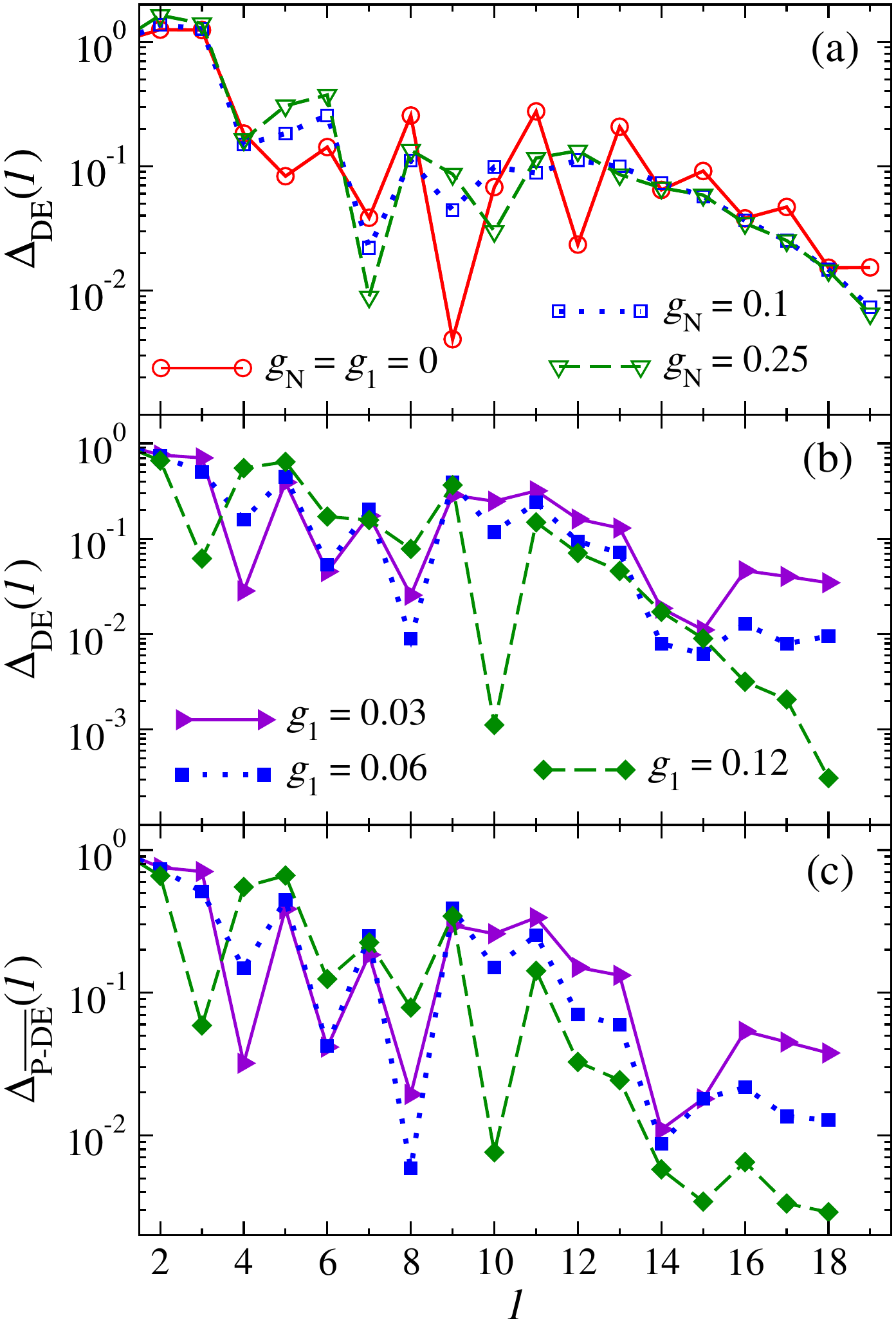}
  \caption{Convergence of the NLCE calculations for the expectation value of $\hat{U}$ in the DE following the quenches: (a) $\hat{H}_I\rightarrow \hat{H}_{g_N}$ and (b) $\hat{H}_I\rightarrow \hat{H}_{g_1}$. (c) Convergence of the $\overline{\text{P-DE}}$ predictions for the quench $\hat{H}_I\rightarrow \hat{H}_{g_1}$ [same quench as in (b)]. The initial state has $T_I=10$ and $\mu_I=1.5$ ($n_I=0.47$). The convergence error at the $l^{\text{th}}$-order is quantified by the relative deviation from the thermal equilibrium values, given by $\Delta_{\text{DE}}(l)$ for the DE and $\Delta_{\overline{\text{P-DE}}}(l)$ for the $\overline{\text{P-DE}}$; see Eq.~\eqref{eq:delta_DE_nlce}.}  
\label{Fig:DE_conv}
\end{figure}

With ``exact'' thermal equilibrium results given by $u^{15}_{{\text{GE}}}$ and $\bar{u}^{15}_{{\text{P-GE}}}$, we can study how the DE and the $\overline{\text{P-DE}}$ converge with increasing the order of the NLCE. We assume that the converged DE calculation agrees with the exact GE one, and quantify the convergence errors using the relative deviations 
\begin{equation}
\Delta_{\text{DE}}(l)=\left|\dfrac{u^{l}_{{\text{DE}}}-u^{15}_{{\text{GE}}}}{u^{15}_{{\text{GE}}}}\right|,\  \Delta_{\overline{\text{P-DE}}}(l)=\left|\dfrac{\bar{u}^{l}_{{\text{P-DE}}}-\bar{u}^{15}_{{\text{P-GE}}}}{\bar{u}^{15}_{{\text{P-GE}}}}\right|, \label{eq:delta_DE_nlce}
\end{equation}
where $u^{l}_{{\text{DE}}}$ and $\bar{u}^{l}_{{\text{P-DE}}}$ are the $l^{\text{th}}$-order NLCE results for the expectation value of $\hat{U}$ in DE and $\overline{\text{P-DE}}$, respectively. Figure~\ref{Fig:DE_conv}(a) shows $\Delta_{\text{DE}}(l)$ for the quench $\hat{H}_I\rightarrow\hat{H}_{g_N}$, while Figs.~\ref{Fig:DE_conv}(b) and~\ref{Fig:DE_conv}(c) show $\Delta_{\text{DE}}(l)$ and  $\Delta_{\overline{\text{P-DE}}}(l)$, respectively, for the quench $\hat{H}_I\rightarrow\hat{H}_{g_1}$.
One can see that $u^{l}_{{\text{DE}}}$ and $\bar{u}^{l}_{{\text{P-DE}}}$ converge towards the thermal equilibrium predictions as $l$ increases (as expected). However the rate of convergence is slower compared to the thermal equilibrium counterparts in Fig.~\ref{Fig:GE_conv}. This is why we push the NLCE calculations for the DE (and also for the dynamics) to the highest orders that can be computed (we evaluate 19 NLCE orders for quenches conserving $\hat{N}$ and 18 NLCE orders in others). 

Note that in Fig.~\ref{Fig:DE_conv}(a) (for the quench that does not support the prethermalization scenario) all values of $g_N\ge0$ show similar convergence of the DE results towards the GE ones. On the other hand, in Figs.~\ref{Fig:DE_conv}(b) and~\ref{Fig:DE_conv}(c) (for the quench that supports prethermalization and thermalization), the convergence worsens as one lowers the value of $g_1$. Because of this, the NLCE for $g_1=0$ [see Fig.~\ref{Fig:DE_conv}(a)] is better converged than the one for $g_1=0.03$. This is similar to the behavior of the NLCE for the DE near an integrable point~\cite{rigol2014quantum}. It shows that the NLCE requires larger clusters to be able to achieve convergence for the DE when $g_1\rightarrow 0$ which, in the thermodynamic limit, is $\mathcal{O}(1)$ different from the prethermal result predicted by the DE (the result for $g_1=0$).   

\section{Difference between the \\ GE and $\overline{\text{P-GE}}$ predictions}\label{appb}

\begin{figure}[!t]
\includegraphics[width=0.985\linewidth]{ 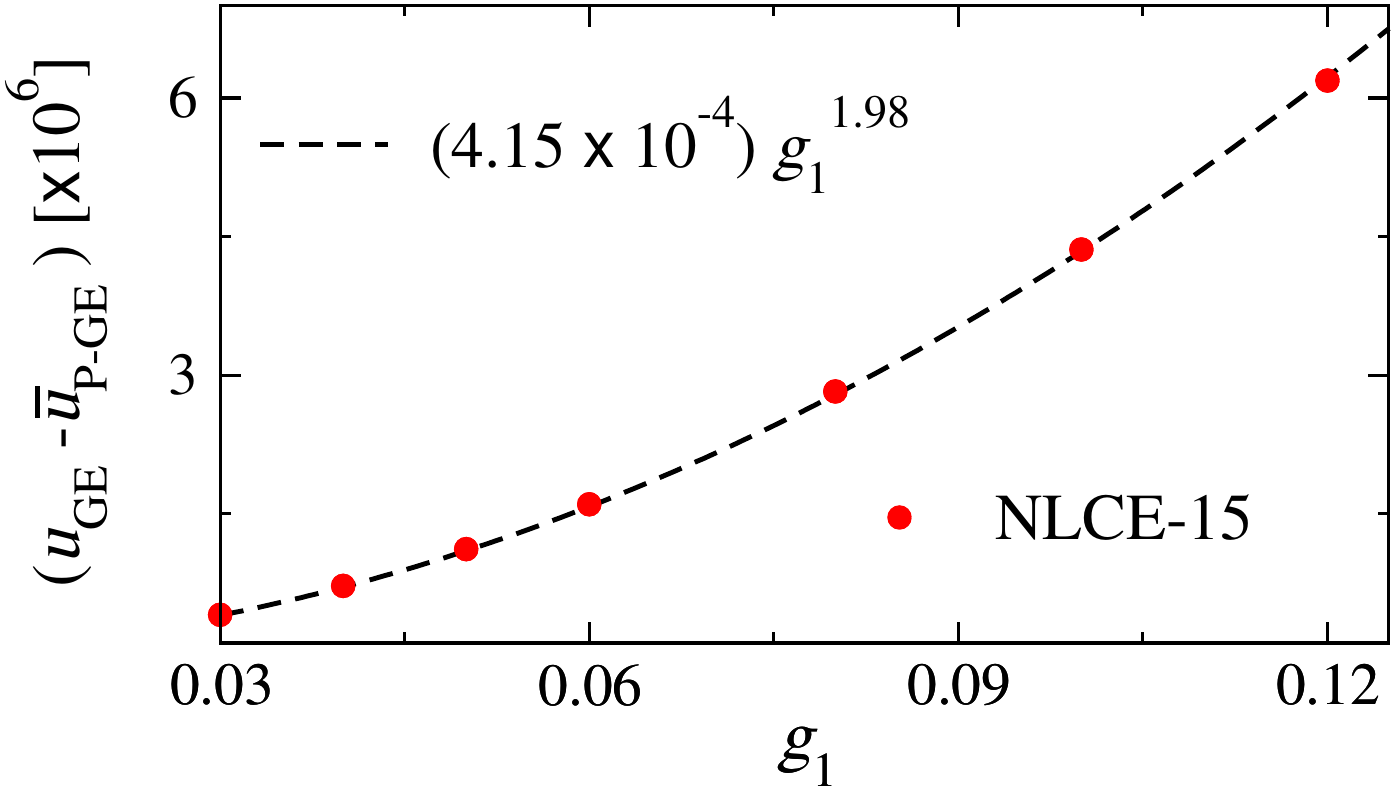}
  \caption{Difference between the expectation value of $\hat{U}$ at long times of the dynamics $(u^{ }_{{\text{GE}}})$ and at long times of the projected dynamics $(\bar{u}^{ }_{{\text{P-GE}}})$, for the quench $\hat{H}_I\rightarrow \hat{H}_{g_1}$ with $0.03\le g_1\le 0.12$, and $n_I=0.47$. The results for $u^{ }_{{\text{GE}}}$ and $\bar{u}^{ }_{{\text{P-GE}}}$ are computed to machine precision using the 15$^{\text{th}}$-order NLCE (NLCE-15). A power-law fit $\alpha g_1^{\beta}$ (dashed line) shows that $(u^{ }_{{\text{GE}}}-\bar{u}^{ }_{{\text{P-GE}}})\propto g_1^2$.}  
\label{Fig:GE_PGE}
\end{figure}

For the quench that leads to a two-step relaxation dynamics ($\hat{H}_I\rightarrow\hat{H}_{g_1}$ with $n_I=0.47$), Fig.~\ref{Fig:Pert_dynamics}(c) shows that the dynamics of the expectation value of $\hat{U}$ $[u(\tau)]$ after prethermalization is closely followed by the projected dynamics $[u^{ }_{{\text{P-DE}}}(\tau)]$. However, a small $\mathcal{O}(g)$ discrepancy is expected to exist in general between the two~\cite{mallayya2019prethermalization}. That difference is most accurately captured by the difference between the long-time thermal equilibrium results $(u^{ }_{{\text{GE}}}-\bar{u}^{ }_{{\text{P-GE}}})$. We compute the latter exactly (within machine precision) using the 15$^\text{th}$-order NLCE. Figure~\ref{Fig:GE_PGE} shows $(u^{ }_{{\text{GE}}}-\bar{u}^{ }_{{\text{P-GE}}})$ plotted vs $g_1$, as well as a power-law fit that makes apparent that the leading order correction is $\propto (g_1)^2$. The vanishing of the $(g_1)^1$ correction term was explained in Ref~\cite{mallayya2019prethermalization} to be a result of the perturbation $\hat{V}_1$ having no terms that conserve $\hat{N}$.

Comparing Figs.~\ref{Fig:O_GE_vs_g} and~\ref{Fig:GE_PGE}, one can see that the relative magnitude of the correction terms is small $(\lesssim 10^{-3})$ for the values of $g_1$ considered in this work. This is why we can accurately describe the slow relaxation regime of $\hat{U}$ using the projected dynamics. 

\section{Relaxation rates from FGR}\label{appc}

Starting from the drift equation~\eqref{fermi_rate}, here we derive the relaxation rate given in Eq.~\eqref{eq:FGR_rate_N}. After block diagonalizing into symmetry sectors $s$ of $\hat{H}_0$ and $\hat N$, the FGR drift equation for $\dot{n}(\tau)\equiv dn/d\tau$ can be written as
\begin{eqnarray}
\dot{n}(\tau)&=&\dfrac{2\pi g_1^2}{L}\sum_{s}\sum_{i,j\in s} \delta(E^0_j-E^0_i) \left(N_j-N_i\right)\nonumber\\&&\times|\bra{E_j^{ N_j}}\hat{V}\ket{E_i^{N_i}}|^2P^0_i(\tau),
\end{eqnarray}
where $\ket{E_i^{N_i}}$ $\left(\ket{E_j^{N_j}}\right)$ are simultaneous eigenkets of $\hat{H}_0$ and $\hat{N}$ within the sector $s$ (associated in our case to space symmetries), with energy $E^0_i$ $(E^0_j)$, and particle number $N_i$ $(N_j)$. Since $\hat{H}_0$ is nonintegrable, $P^0_i(\tau)$ can be replaced by the diagonal matrix elements of $\hat{\rho}^{ }_{_{\text{P-GE}}}(\tau)$ [Eq.~\eqref{eq:rho_Proj_GE}] given by
\begin{equation}
P^0_i(\tau) =  e^{-\beta(\tau) \left[E^0_i+\mu(\tau)N_i\right]} / Z(\tau),
\end{equation}
where $\beta(\tau)=[T(\tau)]^{-1}$ and $Z(\tau)$ is the partition function of the P-GE.  

The perturbation $\hat{V}$ can be split as $\hat{V}=\sum_{\eta}\hat{\mathcal{V}}_{\eta}$, where $\hat{\mathcal{V}}_\eta$ changes the particle number by $\pm \eta$, e.g., $\eta=1$ and 2 for $\hat{V}_1$ [Eq.~\eqref{H1}]. The different $\hat{\mathcal{V}}_{\eta}$ contribute independently to $\dot{n}(\tau)$ to give $\dot{n}(\tau)=\sum_{\eta}\dot{n}_{\eta}(\tau)$, where
\begin{eqnarray*}
\dot{n}_{\eta}(\tau) &=& \dfrac{2\pi g_1^2\eta} {Z(\tau)L} \sum_{s} \sum_{i,j\in s} \delta(E^0_j-E^0_i) e^{-\beta(\tau) \left[E^0_i+\mu(\tau)N_i\right]}\nonumber\\&&\times \left(|\bra{E^{N_i+\eta}_j} \hat{\mathcal{V}}_{\eta} \ket{E^{N_i}_i}|^2 - |\bra{E_j^{ N_i-\eta}} \hat{\mathcal{V}}_{\eta} \ket{E_i^{N_i}}|^2\right).\label{eq:FGR_discrete}
\end{eqnarray*}

Let us assume that the squared magnitude of the matrix elements of $\hat{\mathcal{V}}_\eta$ has no fine-tuned structure within the sector $s$, so that we can define a meaningful coarse-grained
\begin{equation}
F^s_{N,\eta}(E_0)=\text{Avg}_{\Delta E}\left(|\bra{E^{N+\eta}_j}\hat{\mathcal{V}}_{\eta}\ket{E^{N}_i}|^2\right),
\end{equation}
where $\Delta E$ is a small energy window and $E_i^0,\, E_j^0 \in (E_0-\Delta E/2, E_0+\Delta E/2)$. Using $F^s_{N,\eta}(E_0)$, we can replace sums over eigenstates in Eq.~\eqref{eq:FGR_discrete} with integrals over the energy to obtain
\begin{eqnarray}
\dot{n}_{\eta}(\tau)&=&\dfrac{2\pi g_1^2\eta}{Z(\tau)L}\sum_{s}\sum_{N=0}^{L-\eta}\int dE_0  D^{s}_{N+\eta}(E_0)D^{s}_{N}(E_0)e^{-\beta(\tau)E_0}\nonumber\\&&\times\left(e^{-\beta(\tau)\mu(\tau)N}-e^{-\beta(\tau)\mu(\tau)[N+\eta]}\right) F^s_{N,\eta}(E_0),\ \label{eq:FGR_coarse}
\end{eqnarray}
where $D^{s}_{N}(E_0)$ is the density of states of $\hat{H}_0$ at energy $E_0$, particle number $N$, in sector $s$. 

Let us define a rate $\Gamma(\tau)=\sum_{\eta}\Gamma_{\eta}(\tau)$ as 
\begin{eqnarray}
\Gamma_{\eta}(\tau)=-\dfrac{\dot{n}_{\eta}(\tau)}{n(\tau)-n_{\text{GE}}},\label{eq:gamma_tau}
\end{eqnarray}
where $n_\text{GE}$ is the final equilibrium value of the site occupation. After sufficiently long time, when $|n(\tau)-n_{\text{GE}}|\ll 1$, one has that $\mu(\tau)\ll1$, and there is a well defined $\tau$-independent rate $\Gamma_{\eta}$ as shown below.

Since the energy with respect to $\hat{H}_0$ is approximately constant, when $\mu(\tau)$ is sufficiently small (which is the case when $\tau\rightarrow\infty$), $\beta(\tau)$ is also approximately constant: $\beta(\tau)=\bar{\beta}+\mathcal{O}(g_1)$, where $\bar{\beta}^{-1}=\bar{T}$ is the temperature of the $\overline{\text{P-GE}}$ [Eq.~\eqref{eq:proj_finalT_bar}] in the limit $g_1\rightarrow 0$. Then, to leading order in $\mu(\tau)$, $n(\tau)-n_{\text{GE}}$ is given by
\begin{eqnarray}
n(\tau)-n_{\text{GE}}\approx -\dfrac{\bar{\beta}\mu(\tau)\text{Tr}[\hat{N}^2e^{-\bar{\beta}\hat{H}_0}]}{\text{Tr}[e^{-\bar{\beta}\hat{H}_0}]L}. \label{eq:FGR_deno}
\end{eqnarray}

Expanding Eq.~\eqref{eq:FGR_coarse} to leading order in $\mu(\tau)$ gives 
\begin{eqnarray}
\dot{n}_{\eta}(\tau)=-&&\dfrac{2\pi g_1^2\eta^2\bar{\beta}\mu(\tau)}{\text{Tr}[e^{-\bar{\beta}\hat{H}_0}] L}\sum_{s}\sum_{N=0}^{L-\eta}\int dE_0 e^{-\bar{\beta}E_0}\nonumber\\
&&\times D^{s}_{N+\eta}(E_0)D^{s}_{N}(E_0) F^s_{N,\eta}(E_0).\label{eq:FGR_num}
\end{eqnarray}

Substituting Eqs.~\eqref{eq:FGR_deno} and~\eqref{eq:FGR_num} in Eq.~\eqref{eq:gamma_tau} gives the $\tau$-independent rate $\Gamma_{\eta}$ in Eq.~\eqref{eq:FGR_rate_N}.

\section{Numerical evaluation of $\Gamma$ [Eq.~\eqref{eq:FGR_rate_N}]}\label{appd}

We evaluate Eq.~\eqref{eq:FGR_rate_N} numerically using exact diagonalization of chains with $L$ sites and periodic boundary conditions. We use an energy window $\Delta E$ to carry out a coarse graining, by binning the spectrum of $\hat{H}_0$ in each $N$ and $s$ sector. Each bin $\alpha$, with energy $E_\alpha$, includes all eigenkets $\ket{E_i^{N}}$ with energies $E^0_i \in (E_{\alpha}-\Delta E/2, E_{\alpha}+\Delta E/2)$ in the $N$ and $s$ sectors. The density of states at energy $E_\alpha$ is computed as $D^{s}_{N}(E_\alpha) = m_\alpha/\Delta E$, where $m_\alpha$ is the number of energy eigenstates in bin $\alpha$. The coarse-grained function $F^s_{N,\eta}(E_\alpha)$ [Eq.~\eqref{eq:FGR_coarse}] at $E_\alpha$ is evaluated with the energies $E_i^0$ and $E_j^0$ in the bin $\alpha$ for the $N$ and $s$ sector. $\bar{\beta}$ in Eq.~\eqref{eq:FGR_rate_N} is the inverse temperature of $\overline{\text{P-GE}}$ [Eq.~\eqref{eq:proj_finalT_bar}]. For $g_1=0.03$, $\bar{T}\approx 14$, which is a good approximation for the limit $g_1\rightarrow 0$.

\begin{figure}[!b]
\includegraphics[width=0.99\linewidth]{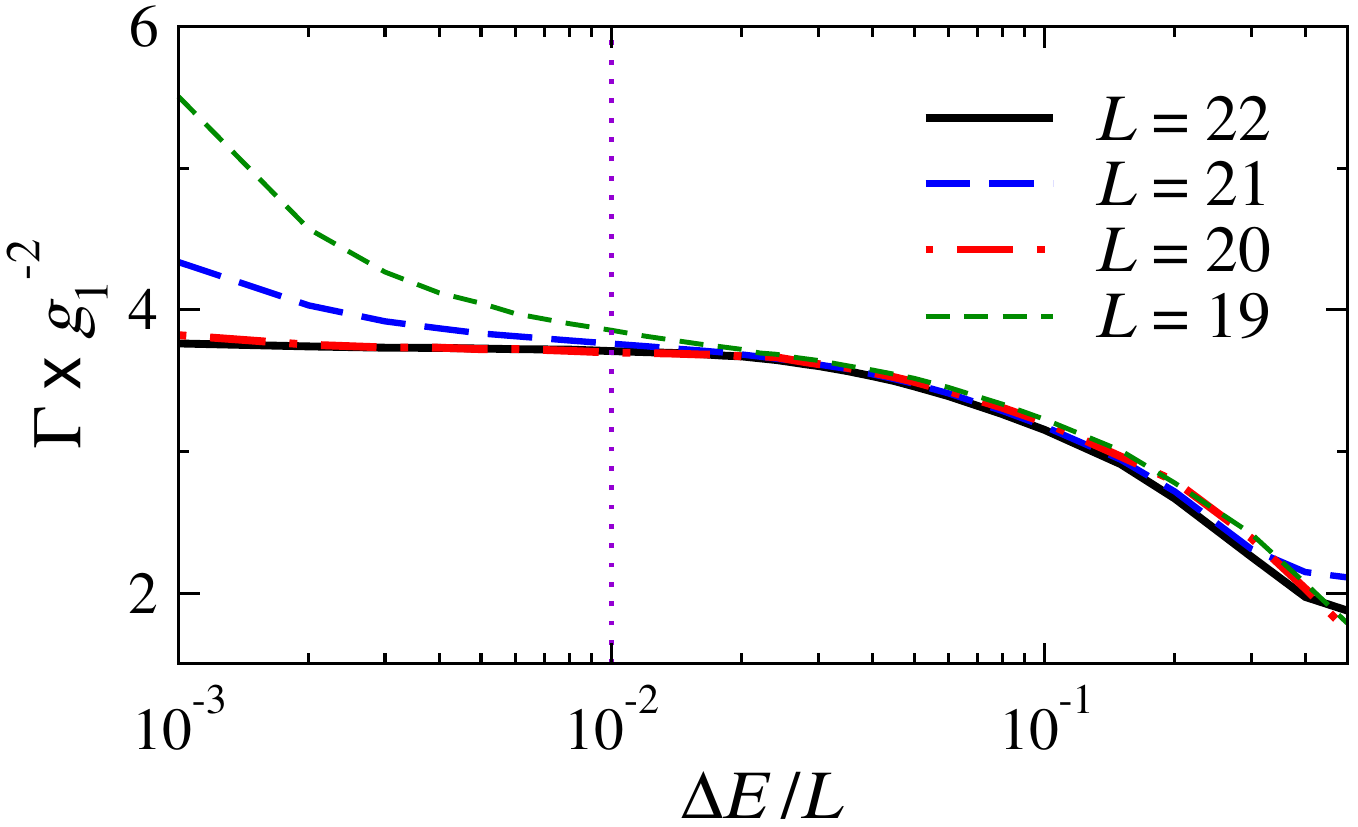}
\vspace{-0.4cm}
  \caption{FGR rate $\Gamma$ estimated evaluating Eq.~\eqref{eq:FGR_rate_N} numerically using ED for different coarse graining energy windows $\Delta E$ and system sizes $L$. The results for $\Gamma$ exhibit a robust regime, nearly independent of the choice of $\Delta E$ and system size, below $\Delta E/L= 10^{-2}$ (vertical dotted line) for $L=22$ and 20. For $L=22$, averaging over $\Delta E/L=\{0.001, 0.002, \dots 0.006, 0.008, 0.01\}$ (eight values) gives $\Gamma/g_1^2\approx 3.73\pm 0.02$, which is the value reported in the main text.}  
\label{Fig:FGR_vs_deltaE}
\end{figure}

In Fig.~\ref{Fig:FGR_vs_deltaE}, we show $\Gamma/g_1^2$ evaluated for different values of $\Delta E$ and $L$. For  $\Delta E/L\lesssim 10^{-2}$ in chains with an even number of sites ($L=22$ and $L=20$), we identify a robust regime in which $\Gamma$ is nearly independent of the choice of $\Delta E$ and finite-size effects are small. For chains with an odd number of sites ($L=21$ and 19), we find that there is a stronger dependence of $\Delta E$ and stronger finite-size effects. Nevertheless, with increasing $L$, the results in those chains approach the ones obtained in even chains. For the FGR results reported in the main text (Figs.~\ref{Fig:Rate_NLCE} and~\ref{Fig:Rate_ED}), we estimate $\Gamma/g_1^2$ with $L=22$, averaged over $\Delta E/L=\{0.001, 0.002, \dots 0.006, 0.008, 0.01\}$ (eight values) giving $\Gamma/g_1^2\approx 3.73\pm 0.02$.

\vspace*{-0.3cm}

\bibliography{Reference}

\end{document}